\documentclass[12pt,preprint,a4paper]{article}

\usepackage{amsmath}
\usepackage{amssymb}
\usepackage{mathrsfs}
\usepackage{graphicx,subcaption}
\usepackage{dsfont}
\usepackage{cite} %for nice listed citations
\usepackage[colorlinks=true,linkcolor=black,citecolor=black,urlcolor=black]{hyperref}
\usepackage{enumerate}
\usepackage{epstopdf}
\usepackage{xcolor}
\usepackage[utf8]{inputenc}

\numberwithin{equation}{section}

\def\half{\frac{1}{2}}
\def\eq#1 { \begin{equation} #1 \end{equation} }

\def\sl2r{SL(2,\mathbb{R})}

\newcommand{\lsim}{\mathrel{\hbox{\rlap{\lower.55ex \hbox{$\sim$}} \kern-.3em \raise.4ex \hbox{$<$}}}}
\newcommand{\gsim}{\mathrel{\hbox{\rlap{\lower.55ex \hbox{$\sim$}} \kern-.3em \raise.4ex \hbox{$>$}}}}

 \newcommand{\be}{\begin{equation}}
\newcommand{\ee}{\end{equation}}

\usepackage{bm}

\newcommand{\lb}{\left}
\newcommand{\rb}{\right}

\newcommand{\mc}{\mathcal}

\newcommand{\mf}{\mathfrak}
\newcommand{\bb}{\mathbb}

\newcommand{\df}[1]{\boldsymbol{#1}}

\newcommand{\gen}[1]{\tt{#1}}

\newcommand{\vd}[1]{#1}

\newcommand{\NR}{\overset{NR}{\mapsto}}

\newcommand{\eqsp}{\hspace{10pt};\hspace{10pt}} %shrtct for space colon space
\newcommand{\defn}{\mathrel{\mathop:}=} %shrtct for definition operator

\makeatletter
\newcommand{\strong}[1]{\@strong{#1}}
\newcommand{\@@strong}[1]{\textbf{\let\@strong\@@@strong#1}}
\newcommand{\@@@strong}[1]{\textnormal{\let\@strong\@@strong#1}}
\let\@strong\@@strong
\makeatother

\begin{document}
%%%%%%%%%%%%%%%%%%%%%%%%%%%%%%%%%%%%%%%%%%%%%%%%%%%%%%%%%%%%%%%%%

\title{\begin{flushright}\vspace{-1in}
       \mbox{\normalsize  EFI-15-14}
       \end{flushright}
       \vskip 20pt
Curved non-relativistic spacetimes, Newtonian gravitation and massive matter}

\date{\today}

\author{
Michael Geracie\thanks{\href{mailto:mgeracie@uchicago.edu}         {mgeracie@uchicago.edu}},~
Kartik Prabhu\thanks{\href{mailto:kartikp@uchicago.edu}         {kartikp@uchicago.edu}},~
   Matthew M. Roberts\thanks{\href{mailto:matthewroberts@uchicago.edu}
     {matthewroberts@uchicago.edu}}
      \\ \\
   {\it\normalsize Kadanoff Center for Theoretical Physics,}\\
   {\it\normalsize Enrico Fermi Institute and Department of Physics}\\
   {\it\normalsize The University of Chicago, Chicago, IL 60637 USA}
} 

\maketitle

\begin{abstract}
There is significant recent work on coupling matter to Newton-Cartan spacetimes with the aim of investigating certain condensed matter phenomena. To this end, one needs to have a completely general spacetime consistent with local non-relativisitic symmetries which supports massive matter fields. In particular, one can not impose a priori restrictions on the geometric data if one wants to analyze matter response to a perturbed geometry. In this paper we construct such a Bargmann spacetime in complete generality without any prior restrictions on the fields specifying the geometry. The resulting spacetime structure includes the familiar Newton-Cartan structure with an additional gauge field which couples to mass.  We illustrate the matter coupling with a few examples. The general spacetime we construct also includes as a special case the covariant description of  Newtonian gravity, which has been thoroughly investigated in previous works. We also show how our Bargmann spacetimes arise from a suitable non-relativistic limit of Lorentzian spacetimes. In a companion paper \cite{GPR-fluids} we use this Bargmann spacetime structure to investigate the details of matter couplings, including the Noether-Ward identities, and transport phenomena and thermodynamics of non-relativistic fluids.
\end{abstract}

\newpage
\tableofcontents

%%=======================================================
\section{Introduction}\label{sec:intro}

Recently there has been a revival of interest in the Newton-Cartan description of non-relativistic spacetimes in the condensed matter literature \cite{Son:2005rv, Hoyos:2011ez, Son:2013, AG, Can:2014ota, Can:2014awa, GS-LLL, GS-transport, GS-Hall-WZW, Gromov:2014vla, GSWW-QHE} where, it has been used with great effect to describe phenomena in the quantum Hall effect and various transport phenomena in condensed matter systems. Newton-Cartan spacetimes are used to describe matter fields and their interaction with general background geometries which are consistent with non-relativistic Galilean invariance. Newton-Cartan geometry also arises in the study of non-relativistic holographic systems, where the boundary theory realizes a ``twistless-torsionful'' Newton-Cartan geometry \cite{Christensen:2013lma, Christensen:2013rfa, Bergshoeff:2014uea, Hartong:2014pma, Hartong:2014oma, Hartong:2015wxa, Hartong:2015zia}. On the other hand, in the gravitational physics literature (see \cite{Cartan-1923, Cartan-1924, DH, Trautman, Kuenzle, Dixon, Kuenzle-NR, Kuchar-Sch, DK, Goenner, DBKP, Ehlers-NR, ABPdR, ABGdR, ABRS} and also Ch.12 of \cite{MTW-book} and Ch.4 of \cite{Mal-book}) Newton-Cartan geometry has been well studied as a diffeomorphism-covariant, geometric way to describe Newtonian gravity. As such, the Newton-Cartan spacetimes considered there belong to a much more restricted class. This divergence of interests has lead to some conflicts in the construction (or at least in the interpretation) of Newton-Cartan spacetimes. One of the aims of this work is to alleviate these conflicts and set a clear stage for describing both Newtonian gravity and matter couplings to non-relativistic spacetimes. To this end, in this paper we will construct the most general spacetime consistent with local Galilean invariance and supporting massive matter fields. In particular, we will construct the geometry with a derivative operator without any a priori restrictions on the torsion, and only later enumerate the invariant conditions that can be imposed to get more restricted spacetimes including those found in the gravitational physics literature. In a companian paper \cite{GPR-fluids}, we provide the analysis of matter fields including fluids, and their Noether-Ward identities and response coefficients coupled to such a general background geometry.\\

Lets recall that a \emph{Newton-Cartan spacetime} consists of a \((d+1)\)-dimensional manifold \(M\), with a corank-\(1\) symmetric tensor field \(h^{\mu\nu}\) of signature $(0,+,\ldots, +)$, a nowhere vanishing 1-form \(n_\mu\) and a derivative operator \(\nabla\) satisfying the compatibility conditions
\be\label{eq:NC-compat}
	n_\mu h^{\mu\nu}=0 \eqsp \nabla_\mu n_\nu = 0 \eqsp \nabla_\mu h^{\nu\lambda} = 0
\ee

The first condition implies that \(n_\mu\) spans the degeneracy direction of \(h^{\mu\nu}\), the the others are reminiscent of the metric compatibility of the derivative opertor in Reimannian (or Lorentzian) geometry. The tensor \(h^{\mu\nu}\) is often called the \emph{Newton-Cartan metric} but we note that since it is degenerate it does not define an actual metric on the manifold \(M\). We'll often refer to the thorough treatment of Newton-Cartan geometry and Newtonian gravity by Malament \cite{Mal-book}, though we will consider the more general case of torsionful spacetimes.

The utility of Newton-Cartan spacetimes in physics comes from identifying \(n_\mu\) as a notion of Galilean clock, and the \(h^{\mu\nu}\) as the spatial metric. We'll defer to  \autoref{sec:barg-geom} to formulate these notions in a precise manner.

Often it is useful to introduce a vector field \(v^\mu\) which denotes the time-direction and satisfies \(v^\mu n_\mu = 1\). Such a choice of vector field can be used to write explicit formulae for instance the Christoffel symbols of a derivative operator (\autoref{eq:Christ}) and also, for writing dynamical laws for matter like the Schr\"odinger field (\autoref{eq:Sch-action-NC}). However, non-relativistic spacetimes can not have a preferred vector field and so we must have invariance under a change of choice of \(v^\mu\), the so called \emph{Milne boosts}
\be\label{eq:Milne-boost}
	v^\mu \mapsto v^\mu + k^\mu \quad\text{where}~ k^\mu n_\mu = 0
\ee
In \cite{Son:2005rv, Hoyos:2011ez, Son:2013, BMM, GSWW-QHE, BMM2, BMM3}, Milne boosts were conflated with certain time-dependent spatial diffeomorphisms acting on tensors fields on \(M\), and the tensor fields were assigned anomalous transformation properties under diffeomorphisms to maintain certain invariance properties. As pointed out in \cite{Jensen:2014aia}, Milne boosts should be rightly considered as additional gauge freedom in the spacetime data, distinct from diffeomorphisms. We show in \autoref{sec:barg-geom} that the freedom in the choice of \(v^\mu\) comes from the ambiguity in choosing a local Galilean frame. All our tensor fields transform as any honest tensor field should under diffeomorphisms.\\

Turning now to the derivative operator \(\nabla\), it is well known (see \cite{DH, Kuenzle} and Prop.4.1.3 of \cite{Mal-book}) that even with the restriction of being torsionless, the Newton-Cartan conditions \autoref{eq:NC-compat} do not uniquely specify \(\nabla\) unlike in the case of Reimannian (or Lorentzian) geometry. We shall work out the details in \autoref{sec:barg-spacetime} but note that the ambiguity is given by a 2-form field \(\Omega_{\mu\nu}\) (called the Coriolis form in \cite{DH-conformalNC}). We'll see that this 2-form encodes the effects of choosing a non-inertial frame using \(v^\mu\) and also under certain restrictions on the spacetime (see \autoref{sec:barg-restrictions}), the Newtonian gravitational potential, so we prefer to call it the \emph{Newton-Coriolis form}. The Newton-Coriolis form can be further restricted by imposing additional constraints on the derivative. For instance, in the torsionless case we can impose the Newtonian condition (see \autoref{eq:Newtonian-cond}) on its curvature. This condition has been used often in relativity literature since it is helpful in describing Newtonian gravity.

In recent condensed matter literature other ways of restricting this freedom in the derivative operator have been proposed. In the torsionless case, in \cite{Son:2013} this was done by demanding that the vector field \(v^\mu\) be curl-free, leading to a derivative operator which is not Milne-invariant\footnote{This isn't surprising since one could always do a Milne boost to another choice of \(v^\mu\) that isn't curl-free.}. In fact we'll show that our Milne-invariant derivative operator reduces to the one used in \cite{Son:2013} when we restrict to the torsionless case and \(v^\mu\) is both curl-free and geodesic.

In \cite{Son:2013, GSWW-QHE} the Newton-Coriolis form was identified with the electromagnetic field tensor. If one then assigns rather strange Milne transformation properties to the electromagnetic gauge field \(A_\mu\) (see Eq.2 of \cite{Son:2013}), the resulting connection is Milne-invariant. But this invariance is spoiled if one adds back torsion into the picture. Moreover, even adding a spin-orbit coupling term to the Schr\"odinger action requires a modification of the Milne transformation of the electomagnetic gauge field (see Eq. 2.32 of \cite{Jensen:2014aia}). This identification of the Newton-Coriolis form with an electromagnetic field is quite unsatisfactory for several reasons. Firstly, the Milne transformation of \(A_\mu\) needed for this to work does not correspond to the way the electromagnetic field should transform (see \autoref{sec:electro} and for instance \cite{BLL}). Indeed, using this anomalous transformation rule one gets spontaneous generation of an electromagnetic field by simply choosing a non-inertial frame! Secondly, while one could take the position that physical spacetimes ought to be torsionless, torsion would be essential in the study of lattices with defects, and of non-relativistic spinor fields. One would also be interested in studying the response of matter fields to perturbations of torsion, even though the background spacetime is torsionless. As we'll show in \autoref{sec:barg-restrictions} the Newton-Coriolis form is closed only under certain restrictions on the spacetime which includes the vanishing of torsion. Thus, one can not identify the Newton-Coriolis form with an electromagnetic field strength in torsionful spacetimes, as was done in \cite{GSWW-QHE}. It would be surprising if there was some essential difficulty in constructing a Newton-Cartan spacetime with a Milne-invariant torsionful derivative operator. In fact, we show that such a derivative operator can be defined quite naturally, and only in the torsionles can it be reduced to the suggestion made in \cite{Son:2013, Jensen:2014aia, Christensen:2013rfa, Christensen:2013lma}, but instead of the electromagnetic gauge field \(A_\mu\) one gets a gauge field for the mass of matter fields. This separation of gauge fields for charge and mass also clarifies the spin-orbit coupling which couples the spin to the usual electromagnetic field and no ad hoc modification of the Milne transformation of the mass gauge field is need to maintain invariance of the Schr\"odinger action. In fact, at the end of Sec.2.4 of \cite{Jensen:2014aia}, Jensen correctly recognizes the mass gauge field but strangely, still uses it in a spin-orbit coupling term in Sec.2.6. with an anomalous Milne-transformation! This separation of the gauge fields for mass and charge also helps in the study of multi-constituent fluids where the constituents can have different charge-to-mass ratios (see \cite{GPR-fluids} for details).\\

Another misunderstood point in recent condensed matter literature is the role of the spacetime in describing gravity. In \cite{AG, Can:2014ota, Can:2014awa}, response of matter fields to deformations of the spatial metric \(h^{\mu\nu}\) has been called gravitational response. Also, the clock form \(n_\mu\) has been often mistakenly accused of encoding gravitational effects. Suppose we pick spacetime coordinates so that \(n_\mu = e^{-\Phi_L}\nabla_\mu t\) for some function \(\Phi_L\) called the \emph{Luttinger potential} (This can only be done in certain restricted spacetimes as discussed in \autoref{sec:barg-restrictions}). In Luttinger's original work \cite{Lutt}, this potential was cleverly used to compute thermal transport coefficients. As pointed out in that work, a varying gravitational potential will produce energy flow and temperature fluctuations. In non-relativistic spacetimes this is no longer true since mass and energy are effectively decoupled, and Newtonian gravitation will produce mass flow but need not produce energy flow or temperature fluctuations. Unfortunately, later uses of the Luttinger potential have often erroneously identified it with the Newtonian gravitational potential in non-relativistic spacetimes\footnote{For instance, the ``gravitational action" in Eq.3.31 of \cite{Jensen:2014aia} includes spatial curvature and a Luttinger potential but does not actually include the effects of Newtonian gravity.}. This misinterpretation might lead one to the incorrect conclusion that a gradient of gravitational potential causes a gradient in the temperature for fluids in equilibrium. While this is true for, say, equilibrium fluid stars in General Relativity where the temperature does ``red/blue shift" in a gravitational field, equilibrium stars in Newtonian gravity have uniform temperature\footnote{We thank Robert M. Wald for pointing out this insightful fact.}. This is also the case for a column of ideal gas in equilibrium under Newtonian gravity where, the temperature remains uniform but the density does fall-off exponentially.

Another troubling issue with this interpretation is that, as we discuss in detail in \autoref{sec:barg-restrictions}, we need \(n_\mu = \nabla_\mu t\) to have a notion of absolute time in non-relativistic spacetimes. If one were to interpret \(\Phi_L\) as the Newtonian gravitational potential, one reaches the incorrect conclusion that Newtonian gravity breaks the non-relativistic notion of absolute time. As has been correctly observed in relativity literature on Newton-Cartan geometry (and as we'll see later), Newtonian gravitational potential actually resides in the connection \(\nabla\), particularly in the Newton-Coriolis form \(\Omega_{\mu\nu}\) (see \cite{Mal-book} for details.). In \cite{GPR-fluids} we compute the associated thermal coefficients and indeed find that the Newtonian potential causes mass density flow while the Luttinger potential causes kinetic energy flow. One can still use perturbations of the spatial metric and the Luttinger potential to obtain Noether-Ward identities and constrain the transport coefficients of fluids, and one could imagine condensed matter systems with an effective Luttinger potential\footnote{As seen in \autoref{sec:barg-restrictions} a non-trivial Luttinger potential still allows the spacetime to have well-defined foliation into spatial hypersurfaces and is appropriately causal.}. We, nevertheless, contend that matter responses to the Luttinger potential or the spatial metric should not be thought of as responses to a Newtonian gravitational potential.\\

With the aim of clarifying these issues, and bridging the divide between the relativity and condensed matter treatments of Newton-Cartan spacetimes, we'll provide a construction of a \emph{Bargmann spacetime}, which is a Newton-Cartan spacetime with an additional gauge-field which couples to the mass of matter fields. We work out the geometry in detail, including the transformations under a local (i.e. gauged) Galilean boosts. We see that choosing a \(v^\mu\) corresponds to picking a local frame basis and Milne boosts are precisely such local Galilean boosts acting on the choice of frame. Using this, we can construct a Milne invariant connection with torsion, and the associated derivative operator, and clarify the role of the Newton-Coriolis form as encoding the non-inertial frame effects and Newtonian gravity. Finally we discuss massive matter fields and their coupling to such a Bargmann spacetime, and for completeness coupling of charged fields to electromagnetism. We leave a discussion of fluids and the computation of Noether-Ward identities for matter to \cite{GPR-fluids}.\\

This paper is organized as follows. In \autoref{sec:barg-spacetime} we introduce the Bargmann group, its Lie algebra, use it to define ``extended coframes" on a manifold, and construct the most general Bargmann spacetime as well as various restricted geometries that might be of interest. We also recover the usual Newtonian gravitational spacetimes as special cases. We further connect our approach with frame bundle reductions and null compactification. \autoref{sec:matter} introduces matter fields of interest and their coupling to the background geometry. For completeness and to clarify certain misconceptions in the recent literature we also elaborate on the coupling of matter to electromagnetism. In \autoref{sec:NR-limit} we show how a Bargmann spacetime arises naturally as a non-relativistic limit of a Lorentzian spacetime, and how the mass gauge field arises in the non-relativistic limit of massive fields, thus providing a solid justification of our approach.\\

We provide a quick overview of the notation used in this paper. For a Lie group \(G\), we refer to its Lie algebra using the corresponding lower-case Gothic letter \(\mf g\). We will use abstract index notation for tensor fields on a manifold as well as fields valued in some vector space. Tensor fields on a manifold will be denoted by abstract indices with lower-cased Greek letters \(\mu,\nu,\ldots\). On manifolds, where we can choose a time function \(t\) and local coordinates \((t,x^i)\) we use \(i,j,\ldots\) as indices in the coordinate basis given by \(x^i\). For abstract indices in vector spaces, we use lower-case Latin letters from the beginning of the alphabet \(a,b,\ldots\) for the vector space \(\bb R^d\), upper-case Latin letters \(A,B,\ldots\) for the vector space \(\bb R^{1+d}\) and upper-case Latin letters from the middle of the alphabet \(I,J,\ldots\) for the vector space \(\bb R^{1+d} \oplus \bb R\). We also find it quite convenient to work with differential forms on a manifold, which we always denote by bold-faced symbols as \(\df \alpha\). To go between the differential forms notation and the abstract index notation, we simply replace the bold-faced letter by the normal letter with abstract indices i.e. for a k-form \(\df\alpha\) we have \(\df\alpha \equiv \alpha_{\mu_1\ldots\mu_k} = \alpha_{[\mu_1\ldots\mu_k]}\) . We follow the sign and numerical factor conventions of Wald \cite{Wald-book} for this translation. In particular, for any k-form \(\df\alpha\) and m-form \(\df\beta\) the \emph{exterior derivative} \(d\) and the \emph{wedge product} \(\wedge\) have the translations
\be\label{eq:df-convention}\begin{split}
	d\df\alpha & \equiv (k+1) \partial_{[\nu}\alpha_{\mu_1\ldots\mu_k]} \\
	\df\alpha \wedge \df\beta & \equiv \frac{(k+m)!}{k! ~m!}\alpha_{[\mu_1\ldots\mu_k}\beta_{\nu_1\ldots\nu_m]}
\end{split}\ee

%%==========================================================

\section{Bargmann group and non-relativistic spacetimes}\label{sec:barg-spacetime}

The flat Galilean spacetime of non-relativistic physics has a symmetry group corresponding to spatial rotations, Galilean boosts, and time and space translations. When massive matter fields are present one needs to extend this group of symmetries by a central element acting as the generator of mass. This can most readily be seen in the case of a particle with mass \(m\). Under a Galilean boost by \(k_i\) the momentum \(p_i\) of the particle transforms to \(p_i \mapsto p_i + m k_i\) while, the mass itself is invariant under any symmetry transformations of the spacetime. Thus, to represent the spacetime symmetries on the one-particle phase space we need a central element corresponding to the mass. The story is similar in the case of a massive Schr\"odinger field and its quantum mechanical description (see for instance \cite{Weinberg-book}).

The central extension of the Galilean group of symmetries by a \(U(1)_M\) symmetry for mass is known as the \emph{Bargmann group} \cite{Barg}. Since we want to construct non-relativistic curved spacetimes with massive matter fields we must start with the Bargmann group as the group of local gauge transformations.

%--------------------------------------------------------

\subsection{Bargmann group and representations}\label{sec:barg-reps}

In a spacetime with \(d\) spatial dimensions the Bargmann group has the structure \cite{Barg}\footnote{In this work we'll only consider the connected component of the Bargmann group.}
\be\label{eq:barg-group}
	Barg(1,d) \defn \lb(SO(d) \ltimes \bb R^{d}\rb) \ltimes \lb( \bb R^{1+d} \otimes U(1)_M \rb)
\ee
where \(\ltimes\) denotes the semi-direct product structure with \(\bb R^{1+d} \otimes U(1)_M\) being the normal subgroup and, the factors have the following physical interpretation: \(SO(d)\) is the group of \emph{spatial rotations}, \(\bb R^d\) denotes the \emph{Galilean boosts}, \(\bb R^{1+d}\) corresponds to translations in spacetime and the \(U(1)_M\) is the central extension corresponding to \emph{mass}. The subgroup of rotations and boosts is the \emph{Galilean group} \(Gal(d) \defn SO(d) \ltimes \bb R^{d} \).

The corresponding Lie algebra is the \emph{Bargmann algebra} with the semi-direct sum structure \(\mf{barg} = \mf{gal} \vdash (\bb R^{1+d} \oplus \bb R )\) where \(\mf{gal} = \mf{so}(d) \vdash \bb R^d\) is Lie algebra of \(Gal(d)\).

At this point we recall the abstract index notation with \(A, B \equiv 0,1,\ldots, d\) on \(\bb R^{1+d}\) and \(a,b \equiv 1,2,\ldots ,d\) in \(\bb R^d\). Using these, the generators of \(\mf{barg}\) can be written as ${{\gen J}^a}_b$, ${\gen K}^a$, \( {\gen P}^A = ({\gen H} ,{\gen P}^a) \) and the central charge \({\gen M}\) which satisfy the commutation relations:
\be\label{eq:barg-comm}\begin{split}
	[ {\gen J}^{ab} , {\gen J}^{cd} ] & = i \left( \delta^{ac} {\gen J}^{bd} - \delta^{ad} {\gen J}^{bc} - \delta^{bc} {\gen J}^{ad} + \delta^{bd} {\gen J}^{ac} \right) \\
	[{\gen J}^{ab} , {\gen P}^c ] &= i \left( \delta^{ac} {\gen P}^b - \delta^{bc} {\gen P}^a \right) \\
	[{\gen J}^{ab} , {\gen K}^c ] &= i \left( \delta^{ac} {\gen K}^b - \delta^{bc} {\gen K}^a \right) \\
	[{\gen P}^a , {\gen K}^b] &= - i\delta^{ab} {\gen M} \\
	[ {\gen H} , {\gen K}^a ] &= - i {\gen P}^a
\end{split}\ee
here \(\delta^{ab} = {\rm diag}(1,1,\ldots,1)\) is the \(SO(d)\)-invariant tensor on \(\bb R^d\).\\

We note here the similarity with the Poincar\'e group and its semi-direct product structure
\be\label{eq:Poin-group}
	Poin(1,d) = SO(1,d) \ltimes \bb R^{1+d}
\ee
with \(\bb R^{1+d}\) being the normal subgroup and the \emph{Lorentz group} \(SO(1,d)\) being a subgroup of spatial rotations and Lorentz boosts. This similarity motivates our construction of a Bargmann spacetime in terms of the ``extended coframes" in analogy to the vielbein formalism for Lorentzian spacetimes. \\

The \emph{fundamental representation} of \(Gal(d)\) on \(\bb F \defn \bb R^{1+d}\) can be written as
\be\label{eq:gal-group-fund}
	{\Lambda^A}_B =
	\begin{pmatrix}
		1 & 0 \\
		- k^a & {\Theta^a}_b \\
	\end{pmatrix}
\ee
where \(k^a \in \bb R^d\) is the parameter for boosts and \({\Theta^a}_b \in SO(d)\) are the usual rotation matrices. 
Thus on \(V^A \in \bb F \) and \(V_A \in {\bb F}^* \) the Galiliean group acts as
\be
	V^A \mapsto {\Lambda^A}_B V^B \eqsp V_A \mapsto V_B{(\Lambda^{-1})^B}_A 
\ee

The group structure of \(Barg(1,d)\) in \autoref{eq:barg-group} also admits an \emph{extended representation} on \(\bb E \defn \bb R^{1+d}\oplus \bb R\) as
\be\label{eq:gal-group-ext}
	{\Lambda^I}_J =
	\begin{pmatrix}
		1 & 0 & 0 \\
		- k^a & {\Theta^a}_b & 0 \\
		-\frac{1}{2} k^2 &  k_c {\Theta^c}_b & 1
	\end{pmatrix}
\ee
This extended representation on \(\bb E\) is completely analogous to the Lorentzian case where \(SO(1,d)\) acts on \(\bb R^{1+d}\), both resulting from the semi-direct product nature of the corresponding groups. We'll make great use of the extended representation to construct both the Bargmann spacetime data and to write \(Gal(d)\)-invariant actions for massive matter fields.\\

Additionally, we have the following \(Gal(d)\)-invariant objects in tensor representations. A \emph{fundamental ``metric"} as corank-\(1\) tensor \(h^{AB}\in \bb F \odot \bb F\), a \emph{volume form} \( \epsilon_{A_0A_1\ldots A_d} \in \bigwedge^{1+d}{\bb F}^* \) (with the sign convention \(\epsilon_{01\ldots d} = 1\)), an \emph{extended metric} \(g^{IJ}\in \bb E \odot \bb E\) and a \emph{clock covector} \(n_A \in {\bb F}^* \). These have the matrix form
\be\label{eq:inv-tensors}
	h^{AB} = 
	\begin{pmatrix}
		0 & 0 \\
		0 & \delta^{ab} 
	\end{pmatrix} \eqsp
	g^{IJ} = \begin{pmatrix}
		0 & 0 & 1 \\
		0 & \delta^{ab} & 0 \\
		1 & 0 & 0
	\end{pmatrix} \eqsp
	n_A = \begin{pmatrix}
		1 & 0
	\end{pmatrix} .
\ee
We note again that \(h^{AB}\) is not invertible and so we do not have an invariant metric on \(\bb F\), while \(g^{IJ}\) is invertible and thus there is a metric (with Lorentzian signature!) on \(\bb E\). Thus, while we can freely use \(g^{IJ}\) and \(g_{IJ}\) (which has the same matrix representation as its inverse) to raise and lower indices on \(\bb E\), we can use \(h^{AB}\) only to raise indices in the fundamental representation. Further, \(n_Ah^{AB} = 0\), which can be seen as the local version of the first Newton-Cartan compatibility condition in \autoref{eq:NC-compat}.\\

The (non-invertible) \(Gal(d)\)-invariant projection \(\Pi : \bb E \to \bb F\) can be written as \(\Pi^A_I = 
	\begin{pmatrix}
		{\delta^A}_B & 0
	\end{pmatrix}
\). We use \(\Pi\) to project from \(\bb E\) to \(\bb F\) and to lift from \(\bb F^*\) to \(\bb E^*\). The projection also intertwines appropriately with the \(Gal(d)\)-representation in the sense
\be\label{eq:proj-intertwine}
	\Pi^A_I{\Lambda^I}_J = {\Lambda^A}_B \Pi^B_J
\ee
For instance, we can lift the clock covector \(n_A\) to define the \emph{extended clock covector} \(n_I \defn n_A \Pi^A_I = \begin{pmatrix}
		1 & 0 & 0 
	\end{pmatrix}\in {\bb E}^*\), and project the extended  metric \(\Pi^A_I\Pi^B_J g^{IJ} = h^{AB}\). Hence forth we use a change of indices to denote a projection or lift with \(\Pi\) and note that due to \autoref{eq:proj-intertwine} there is no conflict with \(Gal(d)\)-invariance in doing so.

%--------------------------------------------------------

\subsection{Bargmann spacetime}\label{sec:barg-geom}

Having elaborated on the structure of the Bargmann group and the representations of interest, we use this structure to construct a \emph{Bargmann spacetime} on a \((1+d)\)-dimensional manifold \(M\) with local (i.e. ``gauged") \(Gal(d)\)-invariance. We proceed in analogy with the relativistic construction of Lorentzian spacetimes, which uses the Poincar\'e algebra. In the Lorentzian case one introduces on \(M\), 1-forms valued in the quotient algebra \(\bb R^{1+d} \cong \mf{poin}/\mf{lor}\) as coframes or \emph{vielbeins} along with an additional 1-form, the \emph{connection} valued in the Lie algebra \(\mf{lor}\) which provides a notion of a covariant derivative (and hence parallel transport).\\

Since we want ``local Galilean invariance" for non-relativitic spacetimes we introduce the \emph{extended coframe} \(\df e^I\) as a 1-form  on \(M\) valued in the normal subgroup \(\bb E \cong \bb R^{1+d} \otimes U(1)_M\) and a \(\mf{gal}\)-valued \emph{connection} 1-form \({\df\omega^I}_J \). We emphasize that we must quotient the full \(Barg(1,d)\) group by the normal subgroup \(\bb R^{1+d} \otimes U(1)_M\) and we can not quotient by just the translations \(\bb R^{1+d}\). This can be interpreted as the fact the spacetime fields which dictate the geometry are massless. If one were to attempt quotienting by just the spacetime translations one would find that the $U(1)_M$ acts nontrivially on the geometric data.  We point out that our construction is a generalization of the ones in \cite{DBKP, DK, ABPdR, ABRS} and we obtain their spacetime under certain invariant restrictions on the torsion and curvature (see \autoref{sec:barg-restrictions}.).

The extended coframe have the decomposition:
\be\label{eq:e-decomp}
	\df e^I = 
	\begin{pmatrix}
		\df n \\ \df e^a \\ \df a
	\end{pmatrix} 
\ee
which defines the \emph{clock form} \(\df n = n_I\df e^I\), the \emph{spatial coframe} \(\df e^a\), and the \emph{mass gauge field}\footnote{We will justify this terminology in \autoref{sec:matter} where we consider matter actions, but here \(\df a\) comes as the \(U(1)_M\) component of the extended coframe.} \(\df a\). We note that the \(\bb F\)-valued coframes \(\df e^A = \Pi^A_I \df e^I\) are the ``real" coframe or vielbeins in the sense that they constitute a set of basis for the cotangent space \(T^*M\). Having pointed out this caveat, we continue the abuse of terminology and call \(\df e^I\) as the ``extended coframe".\\

Similarly, the \(\mf{gal}\)-connection, written either in the fundamental representation as \({\df \omega^A}_B\) or in the extended representation \({\df\omega^I}_J\), can be decomposed as\footnote{Here we use the convention that the connection is a 1-form valued in the Lie algebra, instead of viewing it as 1-form components in a set of bases given by the generators of the Lie algebra.}:
\be\label{eq:omega-decomp}
	{\df\omega^A}_B = 
	\begin{pmatrix}
		0 & 0  \\
		\df\varpi^a & \df\omega^a{}_b 	\end{pmatrix} 
\eqsp
	{\df\omega^I}_J = 
	\begin{pmatrix}
		0 & 0 & 0 \\
		\df\varpi^a & \df\omega^a{}_b & 0 \\
		0 & -\df\varpi_b & 0 
	\end{pmatrix} .
\ee
which defines the \emph{boost connection} \(\df\varpi^a\) and the \emph{spin connection} \({\df\omega^a}_b\). The connection defines a \emph{covariant exterior derivative} \(D\) which for an arbitrary differential form \(\df \alpha^I\) valued in \(\bb E\) takes the form:
\be\label{eq:D-defn}
	D\df \alpha ^I \defn d\df \alpha^I + {\df \omega^I}_J \wedge \df \alpha^J
\ee
Using the Leibniz rule, in the usual way, this defines the action of \(D\) on any differential form valued in \(\bb E\). Note that \(\df\omega^{IJ} = \df\omega^{[IJ]}\) and \(n_I{\df\omega^I}_J = 0\), which respectively lead to \(Dh^{IJ} = 0 \) and \(Dn_I = 0\), which are the local versions of the Newton-Cartan compatibility conditions \autoref{eq:NC-compat}. Similarly \autoref{eq:proj-intertwine} gives \(\Pi^A_I{\df \omega^I}_J = {\df \omega^A}_B \Pi^B_J \) which further leads to \(D\Pi^A_I = 0\), which allows us to freely project or lift indices using \(\Pi\) inside covariant derivatives.\\

The transformation of the coframe and connection under a local boost with parameter \(k^a\) can be computed using \autoref{eq:gal-group-ext} to give:
\begin{subequations}\label{eq:e-omega-trans}\begin{align}
	\begin{pmatrix}
		\df n \\ \df e^a \\ \df a
	\end{pmatrix} \mapsto
		\begin{pmatrix}
		\df n \\ \df e^a - k^a \df n \\ \df a + k_a\df e^a - \half k^2 \df n
	\end{pmatrix} \label{eq:e-trans} \\
	{\df \omega^a}_b \mapsto {\df \omega^a}_b \eqsp \df \varpi^a \mapsto \df \varpi^a + dk^a + {\df \omega^a}_b k^b \label{eq:omega-trans}
\end{align}\end{subequations}

Having introduced the coframe and connection, the \emph{torsion} and \emph{curvature} are defined in the usual way via the \emph{Cartan structure equations}:
\begin{subequations}\label{eq:T-R-defn}\begin{align}
	\df T^I &\defn D\df e^I = d\df e^I + {\df \omega^I}_J \wedge \df e^J \label{eq:T-defn} \\
	{\df R^I}_J &\defn d{\df \omega^I}_J + {\df \omega^I}_K \wedge {\df \omega^K}_J
\end{align}\end{subequations}

The torsion can be decomposed as:
\be\label{eq:T-decomp}
	\df T^I = 
	\begin{pmatrix}
		n_I \df T^I\\ \df T^a \\ \df f
	\end{pmatrix} =
	\begin{pmatrix}
		d\df n \\ d\df e^a + {\df \omega^a}_b \wedge \df e^b + \df\varpi^a \wedge \df n    \\ d\df a - \df \varpi_a \wedge \df e^a
	\end{pmatrix} 
\ee
giving the \emph{clock torsion} \(n_I\df T^I = d\df n\), the \emph{spatial torsion} \(\df T^a\) and the \emph{mass torsion} \(\df f\). We note in particular that \(d\df a\) can \strong{not} be viewed as a curvature, but as a part of torsion and has an additional term depending on the boost connection. This is a direct consequence of the fact that \(\df a\) is not an independent \(U(1)\)-gauge field (like, say an electromagnetic field) but is non-trivially related to the spacetime gauge group through the Bargmann algebra \autoref{eq:barg-comm}.

Since the torsion is also in the extended representation, under local boosts it transforms similar to \autoref{eq:e-trans} i.e.
\be\label{eq:T-trans}
	\begin{pmatrix}
		d\df n \\ \df T^a \\ \df f
	\end{pmatrix}  \mapsto
	 \begin{pmatrix}
		d\df n \\ \df T^a - k^a d\df n \\ \df f + k_a\df T^a - \half k^2 d\df n 
	\end{pmatrix} 
\ee\\

The curvature shows up as the failure of \(D^2\) to vanish i.e. \(D^2\df\alpha^I = {\df R^I}_J \wedge \df\alpha^J\) and can be split into the \emph{boost curvature} \(\df B^a = d\df\varpi^a + {\df \omega^a}_b\wedge \df\varpi^b\) and the \emph{spin curvature} \({\df R^a}_b = d{\df \omega^a}_b + {\df \omega^a}_c \wedge {\df \omega^c}_b\) as:
\be\label{eq:R-decomp}
	{\df R^A}_B = 
	\begin{pmatrix}
		0 & 0  \\
		\df B^a & {\df R^a}_b 
	\end{pmatrix} 
\eqsp
	{\df R^I}_J = 
	\begin{pmatrix}
		0 & 0 & 0 \\
		\df B^a & {\df R^a}_b & 0 \\
		0 & -\df B_b & 0 
	\end{pmatrix} 
\ee
where we have written the curvature in both the fundamental and extended representations. On the individual components a local boost transformation acts as
\be\label{eq:R-trans}
	\df B^a \mapsto \df B^a + {\df R^a}_bk^b \eqsp {\df R^a}_b \mapsto {\df R^a}_b
\ee

The covariant constancy of \(n_A\) and \(h^{AB}\) immediately gives:
\be\label{eq:n-R1}
	n_A {\df R^A}_B = 0 = \df R^{(AB)}
\ee\\

The torsion and curvature also satisfy the \emph{Bianchi identities}
\begin{subequations}\label{eq:Bianchi}\begin{align}
	D\df T^I = D^2\df e^I & = {\df R^I}_J \wedge \df e^J \label{eq:Bianchi1}\\ 
	D{\df R^I}_J & = 0 \label{eq:Bianchi2}
\end{align}\end{subequations}\\

On \(M\), we introduce the \emph{frame} \(e_A^\mu = \begin{pmatrix}v^\mu & e_a^\mu \end{pmatrix}\) as vector fields valued in \(\bb F^*\)  through the relations \(e_A^\mu e^A_\nu = \delta^\mu_\nu\) and \(e_A^\mu e^B_\mu = \delta_A^B\). Under a local boost these transform as
\be\label{eq:inv-e-trans}
	\begin{pmatrix}v^\mu & e_a^\mu \end{pmatrix} \mapsto \begin{pmatrix}v^\mu + k^be_b^\mu & e_a^\mu \end{pmatrix}
\ee
Thus, while the frame transforms covariantly (under the fundamental representation of \(\mf{gal}\)), the vector field \(v^\mu\), being just one component of a covariant object, is not invariant under local boosts\footnote{\(v^\mu\) has sometimes been called an {\ae}ther field, but \(v^\mu\) does not deserve such a misnomer as it very crucially does \emph{not} define a boost-invariant rest frame}.

From \autoref{eq:inv-e-trans} we see that local Galilean boosts by \(k_a\) are precisely the Milne boosts from \autoref{eq:Milne-boost} by a spatial vector field \(k^\mu = k^ae_a^\mu\). It is tempting to think of local boost transformations as being generated by a spatial vector field  as \(v^\mu \mapsto v^\mu + k^\mu\). But we note that, these are \strong{not} diffeomorphisms of \(M\) generated by \(k^\mu\) and in fact, under diffeomorphisms all quantities we have defined transform as tensor fields as they should. To avoid any such confusion, we avoid the terminology of ``Milne boosts" and refer to these transformations as ``local boosts".

It is important to note that one can not define any frame valued in \(\bb E^*\) since the coframe \autoref{eq:e-decomp} are not ``square matrices" and thus have no two-sided inverse. Even if one were tempted to do so by defining \(e_I^\mu ~``\!\!=\!\!"~ \begin{pmatrix}v^\mu & e_a^\mu & b^\mu \end{pmatrix}\) we can see that \(b^\mu\) would be a boost-invariant vector field (a true \emph{{\ae}ther field} or an absolute frame; which has no place in non-relativistic spacetime structure.\footnote{A boost-invariant vector field can be defined, and is useful, when matter fields such as a fluid or a lattice are being considered. In these cases, such a vector field denotes the rest frame of the corresponding matter fields. We will make use of such a rest frame for fluids in \cite{GPR-fluids}.}). Nevertheless we can lift the  frames through the projection \(\Pi\) as \( e_I^\mu \defn \Pi^A_I e_A^\mu = \begin{pmatrix}v^\mu & e_a^\mu & 0 \end{pmatrix} \) but these are only one-sided inverses satisfying \(e_I^\mu e^I_\nu = \delta^\mu_\nu\) but \(e_I^\mu e^J_\mu \neq \delta^J_I\).\\

Having set up this extended coframe formalism we now use it to define tensors on the Bargmann spacetime and connect it to the familiar story of Newton-Cartan spacetimes. We start with a (degenerate, corank 1) ``inverse metric" on \(M\) as
\be\label{eq:metric-defn}
	h^{\mu\nu} \defn h^{AB} e_A^\mu e_B^\nu = \delta^{ab}e_a^\mu e_b^\nu
\ee
which we note is invariant under the local \(Gal(d)\)-transformations. Also, \(n_\mu h^{\mu\nu} = 0 \) which retrieves the first of the Newton-Cartan conditions \autoref{eq:NC-compat}. We'll freely use \(h^{\mu\nu}\) to raise spacetime indices, being aware that this leads to some loss of data.

We can also define a \emph{spacetime volume form} \(\df \varepsilon\) as
\be\label{eq:spacetime-volume}
	\df\varepsilon \defn \frac{1}{(d+1)!} \epsilon_{A_0\ldots A_d}\df e^{A_0} \wedge \ldots \wedge \df e^{A_d}
\ee

Using, the \(\mf{gal}\)-connection \({\df \omega^A}_B\) we can define a \emph{covariant derivative operator} \(\nabla\) on \(M\) using
\be\label{eq:nabla-defn}
	\nabla_\mu e^A_\nu \defn - {{\omega_\mu}^A}_B e^B_\nu
\ee
or equivalently \(\nabla_\mu e_A^\nu = {{\omega_\mu}^B}_A e_B^\nu\). Since both \(n_A\) and \(h^{AB}\) are covariantly constant, this covariant derivative annihilates both the clock form and the inverse metric
\be\label{eq:n-g-compat}
	\nabla_\mu n_\nu = 0 = \nabla_\mu h^{\nu\lambda}
\ee
giving the final two Newton-Cartan compatibility conditions from \autoref{eq:NC-compat}. Thus, we see that the Bargmann spacetime we have constructed is a Newton-Cartan spacetime.\\

The covariant derivative on the vector field \(v^\mu\) gives
\be\label{eq:Dv}
	\nabla_\mu v^\nu = {\varpi_\mu}^a e_a^\nu = {\varpi_\mu}^\nu
\ee
Even though each side of this equation is not boost-invariant, the relation itself is.\\

The covariant derivative operator \(\nabla\) is obviously boost-invariant, being induced by the \(\mf{gal}\)-connection \({\df \omega^A}_B\). But to compare with previous approaches we can define the Christoffel symbols \(\Gamma\) in some coordinate system by \(\nabla = \partial + \Gamma\). To express the Christoffel symbols we need to define certain non-invariant quantities which depend on \(v^\mu\). We start with \({\vd P}^\mu{}_\nu = \delta^\mu_\nu - v^\mu n_\nu = e_a^\mu e^a_\nu\) which projects to vectors orthogonal to \(n_\mu\) and covectors orthogonal to \(v^\mu\). Using this, we can define a \emph{metric relative to \(v^\mu\)} as \({\vd h}_{\mu\nu} = \delta_{ab}e^a_\mu e^a_\nu\) so that \({\vd h}_{\nu\lambda}h^{\mu\lambda} = {\vd P}^\mu{}_\nu\) and \({\vd h}_{\mu\nu}v^\nu = 0\). We'll use this metric \({\vd h}_{\mu\nu}\) to lower spacetime indices again keeping note that this again leads to loss of certain tensor data. Under a local boost by \(k_a\) these transform as
\be\label{eq:P-h-trans}
	{P^\mu}_\nu \mapsto {P^\mu}_\nu - k^\mu n_\nu \eqsp h_{\mu \nu} \mapsto h_{\mu \nu} - n_\mu k_\nu - k_\mu n_\nu + k^2 n_\mu n_\nu 
\ee
where \(k_\mu = k_a e^a_\mu\).

Using these after a tedious but straightforward computation, which we spare the reader, the explicit expression for the Christoffel symbols can be written as
\be\label{eq:Christ}
	{\Gamma^\lambda}_{\mu \nu} = v^\lambda \partial_{(\mu} n_{\nu )} 
	+ \frac{1}{2} h^{\lambda \rho} \left( \partial_\mu {\vd h}_{\nu \rho} + \partial_\nu {\vd h}_{\mu \rho} - \partial_\rho {\vd h}_{\mu \nu} \right) 
	+ \frac{1}{2} \left( {T^\lambda}_{\mu \nu} - \vd T_{\mu \nu}{}^\lambda - \vd T_{\nu \mu}{}^\lambda \right) 
	+ n_{(\mu} {\Omega_{\nu )}}^\lambda
\ee
where we have defined the \emph{spacetime torsion tensor} \({T^\lambda}_{\mu \nu} \defn e_A^\lambda {T^A}_{\mu\nu} = v^\lambda (d\df n)_{\mu\nu} + e_a^\lambda {T^a}_{\mu\nu}\) and the \emph{Newton-Coriolis form} \({\Omega_{\mu\nu}} \defn (\df \varpi_a \wedge \df e^a)_{\mu\nu} = 2{\varpi_{[\mu}}^\lambda~ {\vd h}_{\nu]\lambda}\). None of the individual parts of the above Christoffel symbols are boost-invariant. In fact using \autoref{eq:e-omega-trans} we see that the Newton-Coriolis form transforms as
\be\label{eq:coriolis-trans}
	\df\Omega \mapsto \df\Omega + d\lb( k_a\df e^a -\half k^2\df n  \rb) + \half k^2 d\df n - k_a\df T^a
\ee
Nevertheless one can check that this form of the Christoffel symbols is invariant under local boosts as expected, and \(\nabla\) is a ``Milne-invariant" derivative. We point out that this derivative operator is the same as the one obtained in \cite{BM} using the formalism of \emph{Koszul connections}.\\

As noted before in the Introduction, the Newton-Cartan compatibility conditions \autoref{eq:n-g-compat} by themselves do not determine a unique connection. Our construction in terms of the coframe and connection provides the needed extra data in the form of the boost connection \(\df \varpi^a\) or equivalently through \autoref{eq:Dv} (see also \cite{DBKP, DK, ABPdR}). To interpret this tensor we decompose \( \nabla_\mu v^\nu \) as
\be
	\nabla_\mu v^\nu = {\varpi_\mu}^\nu = n_\mu \alpha^\nu + \half {\sigma_\mu}^\nu + \frac{1}{d}\theta~ {\vd P}^\nu{}_\mu + \half {w_\mu}^\nu 
\ee
into the \emph{acceleration} \(\alpha^\mu\), the \emph{shear} \({\sigma_\mu}^\nu\), the \emph{expansion} \(\theta\) and the \emph{vorticity} \({w_\mu}^\nu\) defined by
\begin{subequations}\label{eq:v-stuff}\begin{align}
	\alpha^\mu & \defn v^\nu \nabla_\nu v^\mu \label{eq:v-acc}\\
	\sigma_{\mu\nu} & \defn 2 {\vd P}^\lambda{}_{(\mu}~ {\vd h}_{\nu)\rho} \lb( \nabla_\lambda v^\rho \rb) - \tfrac{2}{d} \theta~ {\vd h}_{\mu\nu} \label{eq:v-shear}\\
	\theta & \defn \nabla_\mu v^\mu \label{eq:v-exp}\\
	w_{\mu\nu} & \defn  2{\vd P}^\lambda{}_{[\mu}~ {\vd h}_{\nu]\rho} \lb( \nabla_\lambda v^\rho \rb) \label{eq:v-vorticity}
\end{align}\end{subequations}

Thus, the acceleration and vorticity of \(v^\mu\) are precisely the additional data that comes into the Christoffel symbols through \({\Omega_\mu}^\nu =  n_\mu\alpha^\nu + {w_\mu}^\nu\). This fact has been noticed before (see proof of Prop.4.3.4 in \cite{Mal-book}), but its interpretation in terms of a connection for local boosts is new as far as we know\footnote{The boost connection has been used before in the works of \cite{DK, DBKP, ABPdR} but its relation to the acceleration and vorticity of \(v^\mu\) had not been made explicit.}.\\

The Riemann curvature tensor of \(\nabla\) can be obtained from the curvature 2-form using the formula
\be
	{R^\lambda}_{\rho\mu\nu} = ({\df R^A}_B)_{\mu\nu} e_A^\lambda e^B_\rho
\ee
where we note, our index conventions for the Riemann tensor differ from those of Wald \cite{Wald-book}, in particular
\be
	\lb[\nabla_\mu, \nabla_\nu  \rb]\beta_\lambda = - {R^\rho}_{\lambda\mu\nu}\beta_\rho
\ee

The Newton-Cartan compatibility conditions \autoref{eq:n-g-compat} give (also see \autoref{eq:n-R1})
\be\label{eq:n-R2}
	n_\lambda {R^\lambda}_{\rho\mu\nu} = 0 = {R^{(\lambda\rho)}}_{\mu\nu}
\ee
while taking the antisymmetrized derivatives of \(v^\mu\) and \({\vd h}_{\mu\nu}\) respectively give the additional (non-invariant) identities
\begin{subequations}\label{eq:v-g-compat}\begin{align}
	v^\rho {R^\lambda}_{\rho\mu\nu} & = - \nabla_{[\mu}{\Omega_{\nu]}}^\lambda - \half {T^\rho}_{\mu\nu}{\Omega_\rho}^\lambda \label{eq:v-compat}\\
	R_{(\lambda\rho)\mu\nu} = h_{\sigma(\lambda}{R^\sigma}_{\rho)\mu\nu} & = - \nabla_{[\mu}\Omega_{\nu](\lambda}n_{\rho)} - \half {T^\sigma}_{\mu\nu}\Omega_{\sigma(\lambda}n_{\rho)} \label{eq:g-lower-compat}
\end{align}\end{subequations}

For completeness, we recall the spacetime form of the Bianchi identities \autoref{eq:Bianchi}
\begin{subequations}\label{eq:spacetime-Bianchi}\begin{align}
	-\nabla_{[\rho}{T^\lambda}_{\mu\nu]} + {T^\sigma}_{[\rho\mu}{T^\lambda}_{\nu]\sigma} & = {R^\lambda}_{[\rho\mu\nu]} \label{eq:spacetime-Bianchi1} \\
	\nabla_{[\mu|}{R^\lambda}_{\rho|\nu\sigma]} & = {T^\eta}_{[\mu\nu|}{R^\lambda}_{\rho\eta|\sigma]} \label{eq:spacetime-Bianchi2}
\end{align}\end{subequations}

We can define the Ricci tensor \(R_{\mu\nu} \defn {R^\lambda}_{\mu\lambda\nu} \) but due to the presence of torsion it is not symmetric and we have
\be\label{eq:Ricci-antisymm}
	R_{[\mu\nu]} = \frac{3}{2}\nabla_{[\mu}{T^\rho}_{\nu]\rho} - \half {T^\rho}_{\lambda\rho} {T^\lambda}_{\mu\nu}
\ee 
Also, the (spatial) Ricci scalar is
\be\label{eq:Ricci-scalar}
	R \defn h^{\mu\nu}R_{\mu\nu}
\ee\\

Armed with this spacetime data, we can now make a clear connection to the usual Newton-Cartan picture. The clock form \(\df n\) defines a notion of time in the following sense. At a point of \(M\), the \(1\)-dimensional space of covectors spanned by the clock form \(\df n\) (which annihilate \(h^{\mu\nu}\)) are called \emph{temporal} or \emph{time-like}, and the \(d\)-dimensional space of vectors \(\xi^\mu\) which annihilate \(\df n\) (\(\xi^\mu n_\mu = 0\)) are called \emph{spatial} or \emph{space-like}. Also, vectors with \(\xi^\mu n_\mu > 0 \) are \emph{future-directed} while those with \(\xi^\mu n_\mu < 0\) are \emph{past-directed}.  We emphasize that there is in general no invariant notion of ``spatial covectors". The \emph{proper time along a curve} \(\gamma\) parameterized by an arbitrary \(\lambda\), with tangent \(T^\mu\) is given by
\be\label{eq:curve-time}
	\tau = \int_\gamma d\lambda~ T^\mu n_\mu = \int_\gamma \df n
\ee
We can define the \emph{normalized tangent} corresponding to a parameterization of the curve with the proper time as \(\xi^\mu = T^\mu (T^\nu n_\nu)^{-1}\). This will be useful in \autoref{sec:mass-part} to write the action for a massive point particle.

The metric \(h^{\mu\nu}\) in \autoref{eq:metric-defn} is the Newton-Cartan metric and it defines a metric on spatial vectors as follows. If \(\xi^\mu\) is space-like (i.e. \(\xi^\mu n_\mu = 0\)) iff there exists a (not unique) \(\eta_\mu\) such that \(\xi^\mu = h^{\mu\nu}\eta_\nu\). Then, \(h^{\mu\nu}\) defines the length of \(\xi^\mu\) by \((h^{\mu\nu}\eta_\mu\eta_\nu)^\half\), where the non-uniqueness of the \(\eta_\mu\) associated to \(\xi^\mu\) does not matter (see Prop.4.1.1 of \cite{Mal-book}). We can see that on spatial vectors this coincides with the metric defined by \(h_{\mu\nu}\) which on account of \autoref{eq:P-h-trans} is also boost-invariant.

The derivative operator \(\nabla\) in \autoref{eq:nabla-defn} is the correct generalization of the Newton-Cartan derivative to the torsionful case. Thus, we can retrieve all of the usual Newton-Cartan formalism using our extended coframe construction, while also getting ``auxilliary data" in the form of \(\df a\) and \(\df f\). Even though these fields arise quite naturally in the extended coframe formalism, they seem quite unmotivated from a Newton-Cartan perspective and one could wonder if they can be avoided completely. In fact, as we show in \autoref{sec:matter}, these fields couple to massive matter fields living on a background Bargmann spacetime and further in \autoref{sec:NR-limit}, that they arise naturally as order \(c^{-2}\) fields from the non-relativistic limit of Lorentzian spacetimes.\\

We also note that \(g_{IJ}\) can be used to construct an \emph{extended ``metric"} \(G_{\mu\nu}\) on \(M\)
\be\label{eq:ext-G}
	G_{\mu\nu} \defn g_{IJ}e^I_\mu e^J_\nu = 2 n_{(\mu}a_{\nu)} + \delta_{ab}e^a_\mu e^b_\nu = 2 n_{(\mu}a_{\nu)} + {\vd h}_{\mu\nu} 
\ee which agrees with the metric obtained in \cite{Jensen:2014aia} through a null compactification of a Lorentzian spacetime and by \cite{DBKP} from a frame bundle reduction. This extended metric is \(Gal(d)\)-invariant and (at points where \(\df a \neq 0\)) has Lorentzian signature. Further, \(G_{\mu\nu} v^\mu v^\nu = 0\), i.e. \(v^\mu\) is a null vector of the extended metric. This will be useful later to construct invariant matter actions as well as in taking non-relativistic limits, and to relate our approach to the null compactification procedure.

%--------------------------------------------------------

\subsection{Bundle reduction and null compactification}\label{sec:bundle-null}

A well known way to construct a Lorentzian spacetime on a manifold \(M\) is to consider the \emph{linear frame bundle} \(FM\) of the the tangent bundle \(TM\). \(FM\) has the structure of a principal \(GL(1+d)\)-bundle with the fibers at a point \(x\in M\) being the space of linear frames \(e_A: \bb R^{1+d} \to T_xM : \alpha^A \mapsto e_A^\mu \alpha^A \). Similarly the dual-bundle \(FM^*\) provides us with linear coframes \(e^A:(\bb R^{1+d})^* \to T^*_xM : \beta_A \mapsto e^A_\mu \beta_A \). Choosing a preferred section \(\eta_{AB} = {\rm diag}(-c^2,\delta_{ab})\) then reduces the frame bundle to a principle \(SO(1,d)\)-bundle and the general \(\mf{gl}(1+d)\)-connection then reduces to the Lorentzian coframe along with a \(\mf{lor}\)-connection.\\

Attempting to perform a similar reduction of \(FM\) to a principal \(Gal(d)\)-bundle, by choosing preferred sections \(n_A\) and \(h^{AB}\), does work in a completely analogous manner and leads to a construction of Newton-Cartan spacetimes. Except in this case one would retrieve the coframe \(\df e^A\) and miss out on the mass gauge field \(\df a\) and its associated torsion \(\df f\). Such a reduction was done in \cite{DBKP,Jensen:2014aia} and the field \(\df a\) was either completely missed or, added as extra data or some artefact of fixing a ``Milne frame". The mass torsion \(\df f\) was completely missed and so were the non-trivial boost transformations associated with it (see \autoref{eq:T-trans}). This should not be surprising given the Bargmann group structure \autoref{eq:barg-group}, the \(U(1)_M\) part of the group is a central extension and hence acts trivially on the rest of the group and so does not show up in the spacetime frame fields.

The trouble with such an approach is that there is no embedding of the Bargmann group into \(GL(d+1)\). But there is such an embedding into \(GL(d+2)\) \cite{DBKP, DK}. Thus to construct a Bargmann spacetime, one must start with a principal \(GL(d+2)\)-bundle or equivalently a principal \(Barg(1,d)\)-bundle and use \(n_A\) and \(h^{AB}\) to reduce it to a \(Gal(d)\)-bundle. The structure of the group \autoref{eq:barg-group}, then naturally gives us the extended coframes \(\df e^I\) valued in \(\bb E\) and the \(\mf{gal}\)-connection \({\df \omega^I}_J\), and the rest of our construction follows. This the bundle reduction construction was performed in \cite{DK, DBKP} for the restricted class of Newtonian spacetimes.

This also connects to the null compactification procedure in \cite{Jensen:2014aia} (also see \cite{DBKP, JN}). One could reduce the \(GL(d+2)\) bundle to a \(Poin(1,d+1)\)-bundle i.e. construct a Lorentzian spacetime with one extra dimension. The Bargmann group structure then allows one to identify the mass generator \(\gen M\) with translations (or reparameterization) along a null direction (usually parameterized by a coordinate \(x^-\)) in the Lorentizian spacetime. Pulling back the Lorentzian data to the space of orbits of this null isometry we retrieve the extended coframes \(\df e^I\) where now the mass gauge field \(\df a\) can be identified with the Lorentzian coframe in the \(x^-\) direction (Or in the terminology of \cite{Jensen:2014aia} \(\df a\) is the ``graviphoton" of the null reduction.). The pullback of the Lorentzian spacetime metric then gives back the extended metric \autoref{eq:ext-G}. We refer the reader to Sec.3.1. of \cite{Jensen:2014aia} for details of this null reduction procedure.

%--------------------------------------------------------

\subsection{Torsion and restrictions of Bargmann spacetime}\label{sec:barg-restrictions}

The Bargmann spacetime constructed above is quite general. We now analyze the possible \emph{Galilean invariant} restrictions one can place to on the fields dictating the geometry. In the most special case we retrieve spacetimes already explored in the relativity literature with Newtonian gravity. We also mention possible generalizations to the torsionful case that might be of relevance to condensed matter situations.\\

We notice that a general Bargmann spacetime does not even have a well defined notion of ``absolute time" as one expects from a reasonable model of non-relativistic spactime. The trouble is the clock form \(\df n\) can be completely arbitrary apart from satisfying the Newton-Cartan compatibility conditions \autoref{eq:NC-compat}. One restriction that can be imposed, is to have a notion of foliation by spatial hypersurfaces that are orthogonal to \(\df n\). This is guaranteed (at least locally\footnote{We will not deal with global topological issues and assume that the manifold \(M\) has suitable topology so that such hypersurfaces exist globally.}) by \emph{Frobenius' condition} \(\df n \wedge d\df n = 0\). From \autoref{eq:e-trans} and \autoref{eq:T-trans}, we can see that this is a \(Gal(d)\)-invariant condition and hence can be freely imposed. In fact, without this restriction we would have acausal behaviour for curves in the following sense (see \cite{Frankel}). If the Frobenius' condition fails to hold at a point \(x \in M\) then, there is an open neighbourhood of \(x\) in which every point maybe reached by a space-like curve (i.e. curves with tangents \(\xi^\mu\) with \(\xi^\mu n_\mu = 0\)). In such a neighbourhood there is no sensible notion of causality since every point is spatial with respect to \(x\). Thus, we think it is very reasonable to impose Frobenius' condition on \(\df n\) everywhere on \(M\) and we call such spacetimes \emph{causal}. Further this condition is necessary to have some notion of any well-posed initial-value problem (or stated in a quantum language, unitary evolution) for fields living on a Bargmann spacetime, since one would like to prescribe data on some ``initial-time surface" determined by the clock form \(\df n\) and evolve it using the relevant equations of motion. In the Lorentzian case, this is analogous to requiring that the spacetime be \emph{globally hyperbolic} in order to have a well-posed initial-value problem for matter fields with hyperbolic equations of motion.

\begin{figure}[h!]
	\centering
	\includegraphics[width=0.65\textwidth]{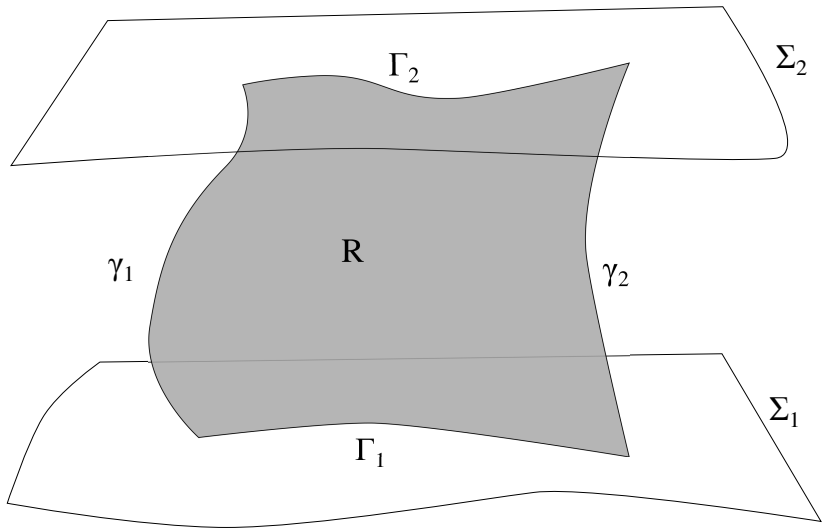}
	\caption{Spacetime diagram with spatial hypersurfaces \(\Sigma_1\), \(\Sigma_2\) and time-like curves \(\gamma_1\) and \(\gamma_2\). The curves \(\Gamma_1\) and \(\Gamma_2\) are arbitrary curves lying entirely in the respective spatial hypersurfaces ending on the intersections of the appropriate $\gamma$ and $\Sigma$. The time measured along \(\gamma_1\) and \(\gamma_2\) differ by the flux of $d\df n$ through $R$.}\label{fig:causal-diag}
\end{figure}

Since \(\df n\) is nowhere vanishing, on such causal Bargmann spacetimes we can write \(\df n = e^{-\Phi_L} dt\) for some functions \(\Phi_L\) and \(t\) and we can treat the \(t = constant\) hypersurfaces \(\Sigma_t\) as the notion of absolute time we seek, and \(\Phi_L\) is the Luttinger potential \cite{Lutt}. We note that while this might give causal evolution it still does not give an absolute time measured by observers along worldlines in spacetime. To see this consider worldlines \(\gamma_1\) and \(\gamma_2\) both beginning at a constant time hypersurface \(\Sigma_1\) and ending on another constant time hypersurface \(\Sigma_2\), then it is easy to see that the difference in the time measured along the wordlines is
\be\label{eq:time-diff}
	\int_{\gamma_2} \df n - \int_{\gamma_1} \df n = \int_R d\df n \neq 0
\ee
where \(R\) is any region of \(M\) bounded by \(\gamma_1\), \(\gamma_2\), and two arbitrary curves \(\Gamma_1\) and \(\Gamma_2\) which lie in \(\Sigma_1\) and \(\Sigma_2\) respectively (see \autoref{fig:causal-diag}). This conclusion holds even if both wordlines start and end at the same points on \(\Sigma_1\) and \(\Sigma_2\), or even if both wordlines are inertial. This lack of absolute time as measured by observers can be attributed to the arbitrary function \(\Phi_L\) which acts as a ``spacetime-dependent unit of time" (very much like the \emph{lapse function} in the relativistic ADM formalism), and hence the time measured along a worldline depends on the history of the wordline. While this is not a problem for relativistic spacetimes which have no observer-independent notion of time, to get the non-relativistic spacetime we know and love we have to further restrict the clock form \(\df n\).\\

As seen from \autoref{eq:time-diff}, the necessary and sufficient condition to get an observer-independent notion of absolute time is the \(Gal(d)\)-invariant condition \(n_I \df T^I = d\df n = 0\) i.e. the clock torsion has to vanish everywhere, therefore \(\df n = dt\) for some \emph{absolute time function} \(t\).\\

\begin{figure}[h!]
	\centering
	\includegraphics[width=.8 \textwidth]{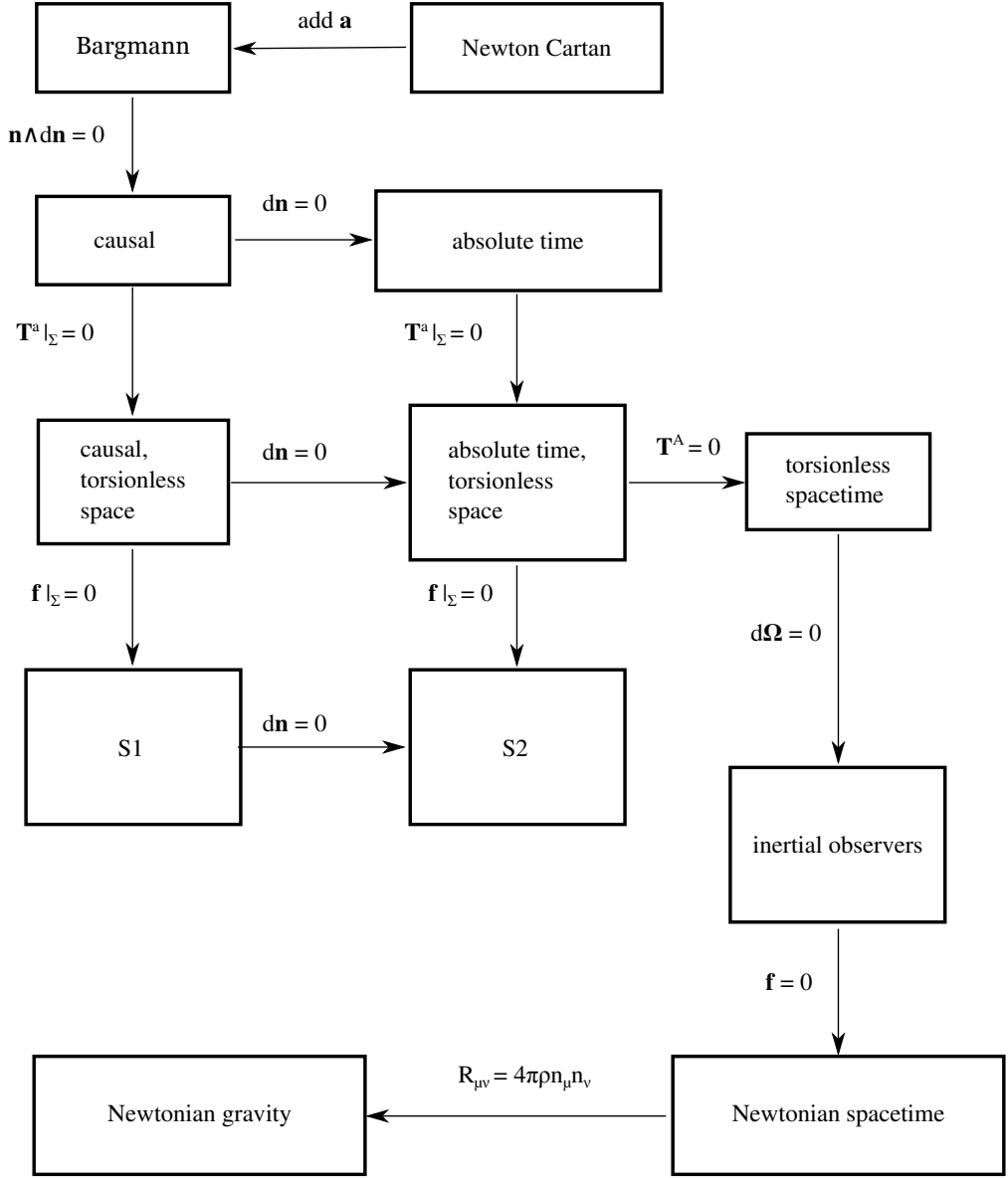}
	\caption{A summary of Bargmann spacetimes and some invariant restrictions that can be imposed on them.}\label{fig:spacetimes}
\end{figure}

On both causal spacetimes and spacetimes with absolute time we see that \(\df n\wedge \df T^a\) is an invariant. Thus, we can impose vanishing of the spatial torsion as \(\df n\wedge \df T^a = 0\). This is easily seen to be equivalent to \(\df T^a\vert_{\Sigma_t} = 0 = T^{\lambda\mu\nu}\). Once we restrict to torsionless space, we can further restrict \(\df f\) to be ``electric" by imposing \(\df n\wedge \df f = 0\), which again is equivalent to \(\df f\vert_{\Sigma_t} = 0\). We do not know of any physical situations that correspond to these restrictions but we note them for completeness and denote them by \(S1\) and \(S2\) in \autoref{fig:spacetimes}.\\

For spacetimes with absolute time we can set the entire spacetime torsion to vanish i.e. \(\df T^A = 0\). Further we can impose a \emph{Newtonian condition} of the form
\be\label{eq:Newtonian-cond}
	{R^{[\lambda}}_{(\rho}{}^{\mu]}{}_{\nu)} = \half {M^{\lambda}}_{(\rho}{}^{\mu}{}_{\nu)} =   0
\ee
where we have written \({M^{\lambda}}_{\rho}{}^{\mu}{}_{\nu} = {R^{\lambda}}_{\rho}{}^{\mu}{}_{\nu} - {R^{\mu}}_{\nu}{}^{\lambda}{}_{\rho}\). We note that while K\"unzle \cite{Kuenzle, Kuenzle-NR} imposes the condition \({R^{[\lambda}}_{(\rho}{}^{\mu]}{}_{\nu)} = 0\), Trautman \cite{Trautman} uses the stronger condition \({M^{\lambda}}_{\rho}{}^{\mu}{}_{\nu} = 0\). Malament \cite{Mal-book} refers to either version as the Newtonian condition since, as we see below, in the torsionless case they are equivalent.

In the torsionless case the Newtonian condition \autoref{eq:Newtonian-cond} can computed by repeated application of \autoref{eq:g-lower-compat} and \autoref{eq:spacetime-Bianchi1} to give 
\be\label{eq:Newtonian-cond-torsionless}
	{M^{\lambda}}_{\rho}{}^{\mu}{}_{\nu} = (d\df\Omega)^{\lambda\mu}{}_{(\rho}n_{\nu)} 
\ee
where we recall that the 2-form \(\df\Omega = \df\varpi_{a}\wedge \df e^a \). Thus, the Newtonian condition implies that \(d\df\Omega = 0\) and hence we can write (in a local coordinate basis \((t,x^i)\))
\be
	\df \Omega = d\lb(\phi +\half\phi_i\phi^i \rb)\wedge \df n  + d\phi_i \wedge dx^i
\ee
where \(\phi_\mu = (\phi_i dx^i)_\mu\) and \(\phi_i\phi^i = h^{\mu\nu}\phi_\mu\phi_\nu\). From this we can recognize the acceleration and vorticity of \(v^\mu\) (see \autoref{eq:v-stuff}) as
\be
	\alpha^\mu = -\half\nabla^\mu\lb(\phi +\half\phi_i\phi^i \rb) \eqsp w_{\mu\nu} = \nabla_{[\mu}\phi_{\nu]}
\ee

In \autoref{sec:NR-limit} we'll show how \(\phi\) and \(\phi_\mu\) arise from the Lorentizian spacetime data (the lapse and shift respectively) in a non-relativistic limit. Once we impose the Newtonian condition on a torsionless spacetime, we see from \autoref{eq:coriolis-trans} that through a local boost transformation we can choose \(\df\Omega = 0\), which corresponds to choosing a frame so that \(v^\mu\) is geodesic and curl-free (see Prop.4.3.3, Prop.4.3.6 and Prop.4.3.7 of \cite{Mal-book}). Then we get the Christoffel symbols used in \cite{Son:2013}
\be\label{eq:Christ-Son}
	{\Gamma^\lambda}_{\mu \nu} = {\hat\Gamma^\lambda}_{\mu \nu} = v^\lambda \partial_{(\mu} n_{\nu )} 
	+ \frac{1}{2} h^{\lambda \rho} \left( \partial_\mu {\vd h}_{\nu \rho} + \partial_\nu {\vd h}_{\mu \rho} - \partial_\rho {\vd h}_{\mu \nu} \right) 
\ee

We also see that the K\"unzle and Trautman versions of the Newtonian condition are equivalent since \({M^{\lambda}}_{\rho}{}^{\mu}{}_{\nu}\) is already symmetric in its lower indices (also see Prop.A.7 of \cite{BM}).\\

In the torsionful case, the analogous computation gives the rather horrendous expression
\be\label{eq:Newtonian-cond-torsionful}\begin{split}
	{M^{\lambda}}_{\rho}{}^{\mu}{}_{\nu} &= (\nabla\df\Omega)^{\lambda\mu}{}_{(\rho}n_{\nu)} + \half\lb( T^{\sigma\lambda\mu}\Omega_{\sigma(\rho} n_{\nu)} - T^{\sigma\mu}{}_\nu {\Omega_\sigma}^\lambda n_\rho + T^{\sigma\lambda}{}_\rho {\Omega_\sigma}^\mu n_\nu  \rb) \\
	&\quad +\frac{3}{2}\lb( {S^\lambda}_\rho{}^\mu{}_\nu - {S^\mu}_\nu{}^\lambda{}_\rho - {S_\rho}^\lambda{}^\mu{}_\nu + {S_\nu}^\mu{}^\lambda{}_\rho\rb)
\end{split}\ee
where \({S^\lambda}_\rho{}^\mu{}_\nu\) is the left-hand-side of \autoref{eq:spacetime-Bianchi1}. We note that Jensen computed a version of the above expression in case \(d\df n \neq 0\) but his expression assumes \(\df T^a = 0\) which, as we have noted, can not be imposed when \(d\df n \neq 0\) in a boost-invariant manner. Evidently, the correct expression \autoref{eq:Newtonian-cond-torsionful} is even more unenlightening than the one obtained by Jensen, and neither the K\"unzle nor the stronger Trautman versions have any reasonable interpretation as far as we can see.\\

We attempt to use our extended coframe formalism to look for a neater way to formulate this condition. For this, we take a look at the first Bianchi identity \autoref{eq:Bianchi1} and split it as
\be\label{eq:first-bianchi}\begin{split}
D\df T^I & = {\df R^I}_J \wedge \df e^J \\
\implies 
	\begin{pmatrix}
		d^2\df n \\
		d\df T^a + {\df\omega^a}_b \wedge \df T^b + \df\varpi^a \wedge d\df n \\
		-d(\df\varpi_a \wedge \df e^a) - \df\varpi_a \wedge \df T^a
	\end{pmatrix} & =
	\begin{pmatrix}
		0 \\
		{\df R^a}_b \wedge \df e^b + \df B^a \wedge \df n \\
		-\df B_a \wedge \df e^a
	\end{pmatrix}
\end{split}\ee
The first component is automatically satisfied for all Bargmann spacetimes. When we have a torsionless spacetime \(\df T^A = 0\), the second component implies that \({\df R^a}_b \wedge \df e^b + \df B^a \wedge \df n = 0\) which gives the usual Bianchi identity on the (Riemannian) spatial curvature on each slice \(\Sigma_t\) \(({\df R^a}_b \wedge \df e^b)\vert_{\Sigma_t} = 0\). Then from the last component we see that the torsionless Newtonian condition \autoref{eq:Newtonian-cond} is equivalent to the restriction (also see Prop.A.5 of \cite{BM}):
\be\label{eq:Newtonian-cond-alt}
	d(\df\varpi_a \wedge \df e^a) = \df B_a \wedge \df e^a = 0
\ee

In case we have a torsionful causal spacetime (\(\df n\wedge d\df n = 0\)) or a torsionful spacetime with absolute time (\(d\df n = 0\)), we propose an invariant \emph{spatial Newtonian condition} of the form
\be\label{eq:Newtonian-cond-spatial}
	d(\df\varpi_a \wedge \df e^a)\vert_{\Sigma_t} = (\df B_a \wedge \df e^a)\vert_{\Sigma_t} = 0
\ee

When the Bargmann spacetime is completely unrestricted, the only invariant generalization of the Newtonian condition would be the \emph{strong Newtonian condition}
\be\label{eq:Newtonian-cond-strong}
	D\df T^I = {\df R^I}_J \wedge \df e^J = 0
\ee
Note that, we can \strong{not} set just the last component to vanish as that would not be boost-invariant.

We'd also like to point out a generalization of the Newtonian condition proposed in Def.5.5 of \cite{BM} which we term the \emph{covariantly exact Newtonian condition}. The condition demands that \(\df\Omega\) be covariantly exact i.e. there exists a 1-form \(\df \beta\) such that
\be\label{eq:Newtonian-exact}
	\Omega_{\mu\nu} = 2 \nabla_{[\mu}\beta_{\nu]} \quad \text{i.e.}\quad \df\Omega = D\beta_A \wedge \df e^A \quad\text{where}\quad \df\beta = \beta_A\df e^A
\ee

 As we have discussed, the usual Newtonian condition \autoref{eq:Newtonian-cond} guarantees the existence of a geodesic, curl-free, future-directed timelike vector field at any point of \(M\) i.e. there exist inertial observers (represented by curl-free, geodesics) through any point of \(M\). A reasonable generalization of the Newtonian condition should imply the existence of such inertial observers or at least some generalization thereof. We defer the analysis of the relation between inertial observers and the generalized Newtonian conditions \autoref{eq:Newtonian-cond-spatial}, \autoref{eq:Newtonian-cond-strong} and \autoref{eq:Newtonian-exact} to future work.\\

Finally, to get a \emph{Newtonian spacetime} (which is the case considered in previous relativity literature), we take the spacetime torsion to vanish \(\df T^A = 0\) and then set \(\df f = 0\) i.e. we set the extended torsion to vanish \(\df T^I = 0\). Denoting then \(\tilde{\df f} = d\df a\), we see that the Christoffel symbols \autoref{eq:Christ} take the form given in \cite{Jensen:2014aia, DBKP, Christensen:2013rfa, Christensen:2013lma, ABPdR}
\be\label{eq:Christ-Jensen}
	{\Gamma^\lambda}_{\mu \nu} = {\tilde{\Gamma}^\lambda}{}_{\mu \nu} = v^\lambda \partial_{(\mu} n_{\nu )} 
	+ \frac{1}{2} h^{\lambda \rho} \left( \partial_\mu {\vd h}_{\nu \rho} + \partial_\nu {\vd h}_{\mu \rho} - \partial_\rho {\vd h}_{\mu \nu} \right) 
	+ n_{(\mu} {\tilde f}_{\nu )}{}^\lambda .
\ee
From the above, it is clear that this form can only be obtained when \(\df T^A = 0\) and thus, explains the failure of previous attempts \cite{Jensen:2014aia,  Christensen:2013rfa, Christensen:2013lma} in obtaining a generalization to the complete torsionful case. This can be traced back to the fact that we can not invariantly set \(\df f = 0\) with non-vanishing torsion---and the appropriate generalization to the torsionful case does not involve \(\tilde{\df f} = d\df a\) but the boost connection \(\df \varpi^a\).

We'll show in \autoref{sec:NR-limit} that a Newtonian spacetime in the above sense is precisely the non-relativistic limit of a torsionless Lorentzian spacetime. On a Newtonian spacetime we can impose the \emph{Newtonian gravitational field equations} \(R_{\mu\nu} = (4\pi \rho - \Lambda) n_\mu n_\nu\) to get \emph{Newton-Hooke gravity} or just \(R_{\mu\nu} = 4\pi \rho n_\mu n_\nu\) to get \emph{Newtonian gravity}, where \(\rho\) denotes the mass density of matter and \(\Lambda\) is the Newtonian analog of a \emph{cosmological constant}. In this case we can reproduce the \emph{K\"unzle-Ehlers Recovery Theorem} Prop.4.5.2 of \cite{Mal-book}, and hence the usual description of Newtonian gravity in a covariant formalism. We note that these field equations automatically impose spatial flatness (see \cite{Kuenzle} and Prop.4.1.5 of \cite{Mal-book}), and the gravitational action in Eq.3.31 of \cite{Jensen:2014aia} does not reproduce these field equations.\\

As seen above, contrary to the Lorentzian case, a Bargmann spacetime has many possible invariant restrictions that can be placed on it to get a richer landscape of potential spacetimes. We summarize some of these in the form of a graph in \autoref{fig:spacetimes}. We hope that some of these would be useful as generalization of Newtonian gravity in the torsionful case, or in other condensed matter applications.

\newpage

%%=========================================================

\section{Matter actions and currents}\label{sec:matter}

To justify the presence of the auxilliary fields \(\df a\) and \(\df f\), we now consider the behaviour of massive matter fields living on a background Bargmann spacetime. We will show that the mass of the fields couples directly to \(\df a\) acting as the gauge field for mass. We also show the utility of the extended coframe formalism in constructing manifestly invariant actions. We defer the analysis of matter Noether currents and the Noether-Ward identities to \cite{GPR-fluids}.

%--------------------------------------------------------

\subsection{Massive particle}\label{sec:mass-part}

We start with the action for a worldline coupled to the background Bargmann geometry i.e. a massive point particle. Let the tangent of the worldline \(\gamma\), parametrized by an arbitrary parameter \(\lambda\) be \(T^\mu\). The \emph{proper time} seen by the particle along \(\gamma\) is
\be
	\tau = \int_\gamma \df n = \int_\gamma d\lambda~ T^\mu n_\mu
\ee

If we parameterize the curve using the proper time \(\tau\), the \emph{normalized tangent} to \(\gamma\) is given by \(\xi^\mu \defn T^\mu \lb( T^\mu n_\nu \rb)^{-1}\). Using the extended coframes, we can write the tangent vector as \(\xi^I \defn e^I_\mu\xi^\mu \in \bb E\). Then the action we propose for a particle of mass \(m\) is
\be\label{eq:particle-action}
	\mc S_{\rm particle} \defn \frac{m}{2}\int_\gamma d\tau~ g_{IJ}\xi^I \xi^J
\ee
which is manifestly invariant. To see that this is the same as previously proposed actions we use the extended metric from \autoref{eq:ext-G} and compute
\be\label{eq:particle-action-NC}\begin{split}
	\mc S_{\rm particle} &= \frac{m}{2}\int_\gamma d\tau~  g_{IJ}e^I_\mu e^J_\nu \xi^\mu \xi^\nu = \frac{m}{2}\int_\gamma d\tau~  G_{\mu\nu} \xi^\mu \xi^\nu \\
		& = \frac{m}{2}\int_\gamma d\tau~  {\vd h}_{\mu\nu}\xi^\mu \xi^\nu + m\int_\gamma \df a
\end{split}\ee
which agrees with \cite{Kuenzle, Jensen:2014aia, ABGdR}. In fact, the first term is just the \emph{kinetic energy} of the particle and the last term is the \emph{potential energy} due to interaction with the background Bargmann geometry (in a Newtonian spacetime, this is just the gravitational force and possible Coriolis terms). We'll see that in Newtonian spacetimes this last term corresponds precisely to potential energy arising due to a Newtonian gravitation and a non-inertial choice of frame. This separation of the energy of the particle into kinetic and potential parts, depends on the choice of Galilean frame as shown by the appearance of \({\vd h}_{\mu\nu}\) in the kinetic energy.\\

Even though the final form in \autoref{eq:particle-action-NC} is not manifestly covariant, the form \autoref{eq:particle-action} is. This shows the utility of our use of the extended frame formalism in constructing manifestly invariant actions even though invariance, especially under local boost, is not so apparent in spacetime terms.\\

The wordline equation of motion obtained by varying the action \autoref{eq:particle-action-NC} can be decomposed into spatial and temporal parts by contracting the appropriate indices with \(h^{\mu\nu}\) and \(v^\mu\) respectively. The spatial part gives a geodesic equation,
\be
\xi^\nu \nabla_\nu \xi^\mu =  {f^\mu}_\nu \xi^\nu - {T_{\nu\lambda}}^\mu \xi^\nu \xi^\lambda- \half \left(h_{\rho\lambda}\xi^\rho \xi^\lambda \right) (d\df n)^\mu{}_\nu \xi^\nu
\ee
Thanks to our extended representation, we may rewrite the not-obviously-invariant right hand side in an explicitly invariant manner,
\be
\xi^\nu \nabla_\nu \xi^\mu = u_I \left( \df T^I\right)^\mu{}_\nu \xi^\nu,\quad \text{where}~u^I =\begin{pmatrix} e^A_\mu \xi^\mu \\ -\half \xi^2 \end{pmatrix}.
\ee\footnote{While it is not immediately obvious one can check that $u^I$ defined this way is an element of $\bb E$ and transforms covariantly.}
The temporal part gives a \emph{work-energy equation},
\be
\xi^\mu \nabla_\mu  \left(h_{\nu\lambda}\xi^\nu \xi^\lambda \right)= (v^\mu f_{\nu\mu})\xi^\nu -\half \left(h_{\nu\rho}\xi^\nu \xi^\rho \right) v^\lambda (d\df n)_{\rho\lambda}\xi^\rho + \alpha^\mu h_{\mu\nu}\xi^\nu - \half (\pounds_v h_{\mu\nu} )\xi^\mu \xi^\nu,
\ee
which is of course simply a consequence of the invariance of the action \autoref{eq:particle-action} under reparameterization. On the right-hand-side we can clearly see the effects of the work done by the background torsion in the first two terms. The last two terms represent the effects of a non-inertial choice of frame where we recall the acceleration \(\alpha^\mu\) from \autoref{eq:v-stuff} and note that \(\pounds_v h_{\mu\nu} = \sigma_{\mu\nu} + \tfrac{2}{d}\theta h_{\mu\nu} \). The work-energy equation can also be written invariantly as
\be
\xi^\mu \nabla_\mu \mathcal{L}_{\rm particle} = 0.
\ee

When \(\df T^I \neq 0\) we have extra ``forces'' on the particles due to the geometry, and the resulting Bargmann spacetime violates the equivalence principle. This is to be expected, as from Lorentzian spacetime we know that torsion in the derivative also acts as an external force violating the equivalence principle and we'll show in \autoref{sec:NR-limit} that \(\df f\) arises as a part of the Lorentzian torsion in the non-relativistic limit. In Newtonian spacetimes where \(\df T^I = 0\) we get back the usual geodesic equation for massive particles \cite{Kuenzle}.
%--------------------------------------------------------

\subsection{Schr\"odinger field}
The simplest field on a Bargmann spacetime would be the \emph{Schr\"odinger field} \(\psi\) which is a massive, complex scalar field. Such a field is valued in the trivial representation of \(Gal(d)\) but under a massive representation of \(U(1)_M\) of the Bargmann algebra. On such a field the mass generator \(\gen M\) has the action
\be\label{eq:Sch-mass}
	{\gen M}\psi = im\psi \eqsp {\gen M}\bar\psi = -im\bar\psi
\ee
where \(m\) labels the representation of \(U(1)_M\) with the physical interpretation of the mass of the Schr\"odinger field.

Thus, the natural covariant derivative \(D\) for massive fields is given by adding the mass gauge field \(\df a\) in the usual way
\be
	D_\mu \defn \nabla_\mu - a_\mu \gen M
\ee

Using this, we now wish to write a covariant action for the Schr\"odinger field which includes \(\psi\), \(\bar\psi\) which is at most quadratic in their first derivatives and real. The na\"ive guess \(h^{\mu\nu}D_\mu\bar\psi D_\nu\psi\) does not work since it has no time derivatives and thus, will not give a dynamical equation.

We see that the time derivative can be included if we use \(D_0 = e_0^\mu D_\mu = v^\mu D_\mu\) and to keep things covariant then we must use the full operator \(D_A = e_A^\mu D_\mu\). The correct action turns out to be \cite{Kuchar-Sch, DBKP, DK, Jensen:2014aia}
\be\label{eq:Sch-action-NC}
	\mc S_{\rm Sch} = \int_M \df\varepsilon~ \lb(im \bar\psi v^\mu D_\mu\psi - im v^\mu D_\mu\bar\psi \psi - h^{\mu\nu}D_\mu\bar\psi D_\nu\psi \rb)
\ee
Choosing a flat Galilean spacetime with a choice of frames \(\df n = dt\) and \(v^\mu = (\partial_t)^\mu\) one retrieves the usual textbook massive flat space Schr\"odinger action.

Even though this action is invariant under \(Gal(d)\) and \(U(1)_M\), it is written as a rather strange combination of non-invariant terms, and one needs to carefully check for invariance (particularly under local-boosts). It would be enormously useful if one could write this action in a manifestly invariant way.\\

To do so, we repackage the derivatives and \(\gen M\) into an operator \(D_I\) valued in \(\bb E^*\) as follows
\be
	D_I \defn \begin{pmatrix}D_A & {\gen M}\end{pmatrix} = \begin{pmatrix}e_A^\mu D_\mu & {\gen M}\end{pmatrix}
\ee

Using this, the obvious quadratic action to write down is
\be\label{eq:Sch-action}
	\mc S_{\rm Sch} \defn -\int_M \df\varepsilon~ g^{IJ}D_I\bar\psi D_J\psi
\ee
Expanding the above form, a straightforward computation then gives back \autoref{eq:Sch-action-NC}. This is again an illustration of the utility of the extended frame formalism we have introduced. The Schr\"odinger action takes a particularly simple and manfestly invariant form (identical to a \emph{massless} scalar field in Lorentzian spacetime) when written in terms of the extended frame indices.

One can easily add a term \(V(x)\bar\psi\psi\) in the Lagrangian above where \(V\) is a real function on spacetime \(M\), and one gets the Schr\"odinger equation with an external potential. We discuss possible higher derivative interactions and the Noether currents for a Schr\"odinger field in \cite{GPR-fluids}.\\

The Schr\"odinger equation derived from the action \autoref{eq:Sch-action} is
\be\label{eq:Sch-eqn}
	g^{IJ}D_I D_J\psi = 2im v^\mu D_\mu\psi + h^{\mu\nu}D_\mu D_\nu\psi  = 0
\ee
which is a \emph{second-order parabolic PDE} (see \cite{Evans-book}). On causal spacetimes (\(\df n \wedge d\df n = 0\)) we can then study the initial-value-problem and hence the evolution in time (given by a foliation \(\Sigma_t\) by spatial hypersurfaces orthogonal to \(\df n\)) for a Schr\"odinger field \(\psi\). For general Bargmann spacetimes, which are not causal, there is no reasonable notion of time evolution. One could attempt to choose an arbitrary local coordinate system \((t,x^i)\) and give initial-data on the \(t=constant\) surfaces. However, it is easy to see that since we no longer have \(\df n = e^{-\Phi_L}dt\), a local boost can make \(v^\mu\) tangent to the arbitrarily chosen \(t=constant\) surfaces, and the Schr\"odinger equation is not guaranteed to be parabolic with respect to such an arbitrary notion of time. 

%---------------------------------------------------------

\subsection{Electromagnetism}\label{sec:electro}

We can introduce electromagnetism into the picture in a fairly straightforward way. Now matter fields are charged under an additional \(U(1)_Q\) group which appears as a direct product with the Bargmann group. We introduce the corresponding electromagnetic gauge field \(\df A\) and its curvature \(\df F \defn d\df A\). Both fields \(\df A \) and \(\df F\) are completely invariant under both \(Gal(d)\) (manifest by the absence of any \(Gal\)-indices), and \(U(1)_M\), reflecting the fact that electromagnetic fields are massless.\\

In some choice of local frame \(e_A^\mu\) we can decompose these forms as
\be\begin{split}
	A_A &\defn e_A^\mu A_\mu = \begin{pmatrix}\varphi, A_a\end{pmatrix} \\
	F_{AB} &\defn  e_A^\mu e_B^\nu F_{\mu\nu} = \begin{pmatrix}0 & -E_b \\ E_a & B_{ab}\end{pmatrix}
\end{split}\ee
Here \(\varphi\) and \(A_a\) are the \emph{electric potential} and \emph{vector potential} as observed in the chosen local frame while \(E_a\) and \(B_{ab}\) are the \emph{electric} and \emph{magnetic} fields. Then using \autoref{eq:e-omega-trans} we get the following boost transformations of the electromagnetic data
\be\label{eq:A-F-trans}\begin{split}
	\varphi \mapsto \varphi + k^a A_a &\eqsp A_a \mapsto A_a \\
	E_a \mapsto E_a + k^b B_{ab} &\eqsp B_{ab} \mapsto B_{ab}
\end{split}\ee
which are reminiscent of the boost-transformations of electric and magnetic fields on flat Galilean spacetime (see \cite{BLL}).\\

To add coupling of such an electromagnetic field to charged matter fields, one simply uses the charge derivative operator \(D_\mu = \nabla_\mu - a_\mu {\gen M} - A_{\mu}{\gen Q}\) where \({\gen Q}\) is the generator of electromagnetic charge in \(U(1)_Q\). It is straightforward to see that using this we get the action for a charged, massive Schr\"odinger field coupled to background spacetime and electromagnetism. We see that we can add a covariant coupling of the Schr\"odinger field to $A_\mu$ without ``modifying" the transformation of \(\df a\) under local-boosts. There is similarly no obstruction to introducing spin-orbit coupling.

%%======================================================

\section{Non-relativistic limit of Lorentzian spacetimes}\label{sec:NR-limit}

A pertinent question is whether a Bargmann spacetime can arise naturally as the non-relativisitic limit of some \((1+d)\)-dimensional Lorentzian spacetime. And how does the extra data \(\df a\) and \(\df f\) come in? To answer these, we analyse the non-relativistic limits of Lorentzian spacetimes in the coframe formalism and compare it to the data in a Bargmann spacetime. Non-relativisitic limits of Lorentzian spacetimes were considered in detail in \cite{Kuenzle-NR} (also see \cite{Ehlers-NR} for some example spacetimes.).

As is well-known, we can obtain the Bargmann algebra from the Poincar\'e algebra by taking a group contraction \cite{IW} (also see \cite{ABPdR, Weinberg-book})  with the limit \(c \to \infty\) which we denote by the symbol \(\NR\). To see this we write the Lorentz invariant metric as \(\eta_{AB} = {\rm diag}(-c^2,\delta_{ab})\) and \(\eta^{AB} = {\rm diag}(-\tfrac{1}{c^2},\delta^{ab}) \). With this convention we can take the limit of the Lorentz generators as
\be\label{eq:Lor-gen}
	{{\gen J}^a}_b \NR {{\gen J}^a}_b \eqsp {{\gen J}^a}_0 = {\gen K}^a \NR {\gen K}^a \eqsp {{\gen J}^0}_a = \tfrac{1}{c^2}{\gen K}_a \NR 0
\ee
and the Poincar\'e translations \({\gen P}^A = \begin{pmatrix}{\gen P}^0 & {\gen P}^a\end{pmatrix}\) as
\be\label{eq:Poin-gen}
	{\gen P}^0 = {\gen H} + \frac{\gen M}{c^2} \NR {\gen H} \eqsp {\gen P}^a \NR {\gen P}^a
\ee

Using these scalings we see that \(\mf{poin} \NR \mf{barg}\). In the same vein, we get the contraction of the \(\mf{lor}\)-connection to a \(\mf{gal}\)-connection
\be\label{eq:Lor-conn}
	{\df \omega^A}_B \defn 
	\begin{pmatrix}
		0 & {\df \omega^0}_b   \\
		{\df \omega^a}_0 & {\df \omega^a}_b
	\end{pmatrix}
	= 	\begin{pmatrix}
		0 & \tfrac{1}{c^2}\df \varpi_b  \\
		\df \varpi^a & {\df \omega^a}_b
	\end{pmatrix}
	\NR 	\begin{pmatrix}
		0 & 0 \\
		\df \varpi^a & {\df \omega^a}_b
	\end{pmatrix}
\ee

Similarly we may retrieve the raised spatial metric from \(\eta^{AB} \NR h^{AB}\) while the clock covector \(n_A = (1,0)\) arises from  \( \tfrac 1 {c^2} \eta_{AB} \NR - n_A n_B\). We further see that it transforms as \(n_A {{\gen J}^A}_B = \frac{1}{c^2}K_b \NR 0\). Thus, we retrieve \(n_A\) and \(h^{AB}\) as invariant tensors in the non-relativistic limit.\\

To start with a simple case, we'll show in the following that the non-relativistic limit of a torsionless Lorentzian spacetime is a Newtonian spacetime in the sense of \autoref{sec:barg-restrictions}. Later we describe how to add torsion back in for more general cases. To perform the contraction, we recall the Lorentzian metric of the \emph{ADM decomposition} in General Relativity (see Sec.E.2 of \cite{Wald-book} or Sec.VI.3 of \cite{CB-book}), with a choice of time function \(t\) and local coordinates \((t,x^i)\), given by
\be\label{eq:ADM}\begin{split}
	g &\defn -c^2 N^2 dt \otimes dt + h_{ij}(c N^i dt + dx^i) \otimes (c N^j dt + dx^j)  \\
	g^{-1} &= - \frac{1}{c^2}N^{-2}\lb(\partial_t - c N^i \partial_i\rb) \otimes \lb(\partial_t - cN^j \partial_j \rb) + h^{ij}\partial_i \otimes \partial_j
\end{split}\ee
with the \emph{lapse} \(N\), the \emph{shift} \(N^i\) and the \emph{spatial metric} \(h_{ij}\). The clock form is then \(\df n = dt\) and to get an appropriate non-relativistic limit we should choose the scalings \(N^2 = 1 - \frac{2\phi}{c^2} + O(c^{-3})\), \(N^i = \frac{\phi^i}{c} + O(c^{-2})\) and \(h_{ij} = O(1)\). 

To compute the boost connection for a Newtonian spacetime in the non-relativistic limit we write the Lorentzian metric above in the form \(g = \eta_{AB}\df e^A \otimes \df e^B\) by a choice of Lorentzian coframe and (assuming a torsionless Lorentizian spacetime) compute \( d\df e^0 = - {\df\omega^0}_a \wedge \df e^a \NR -\tfrac{1}{c^2}\df\varpi_a \wedge \df e^a \). Thus, the non-relativistic Newton-Coriolis form would be given by
\be\label{eq:Coriolis-NR}
	-c^2\lb( d\df e^0 \rb) \NR \df\Omega
\ee

Using the local freedom in the choice of Lorentizian vielbeins we can perform this computation in different \(\mf{lor}\)-gauges. We illustrate two possible choices given as follows:
\begin{subequations}\label{eq:lor-choice}\begin{align}
\begin{split}
	\df e^0 = N dt &\eqsp \df e^a = \beta^a_i \lb( c N^i dt + dx^i \rb) \\
	e_0^\mu \equiv N^{-1}\lb(\partial_t - c N^i \partial_i\rb)^\mu &\eqsp e_a^\mu \equiv \beta_a^i (\partial_i)^\mu
\end{split}\label{eq:lor-choice1} \\[20pt]
\begin{split}
	\tilde{\df e}^0 = \lb(N^2 - N_i N^i \rb)^\half dt - \frac{N_i}{c}\lb(N^2 - N_i N^i \rb)^{-\half}dx^i  &\eqsp \tilde{\df e}^a = \beta^a_i dx^i \\
	e_0^\mu \equiv \lb(N^2 - N_i N^i \rb)^{-\half}(\partial_t)^\mu  &\eqsp e_a^\mu \equiv \beta_a^i \lb( - \frac{N_i}{c}  \lb(N^2 - N_i N^i \rb)^{-1} \partial_t + \partial_i \rb)^\mu
\end{split}\label{eq:lor-choice2}
\end{align}\end{subequations}
with \(\delta_{ab}\beta^a_i\beta^b_j = h_{ij}\), \(\delta^{ab}\beta_a^i\beta_b^j = h^{ij}\) and \(\beta_a^i \equiv \lb( \beta^a_i \rb)^{-1} \).

Using the prescribed scalings in the \(c \to \infty\) limit we get
\begin{subequations}\label{eq:NR-choice}\begin{align}
\begin{split}
	\df e^0 \NR dt = \df n &\eqsp \df e^a \NR \beta^a_i\lb( \phi^i \df n + dx^i \rb) \\
	e_0^\mu \NR (\partial_t)^\mu - \phi^i (\partial_i)^\mu = v^\mu &\eqsp e_a^\mu \NR \beta_a^i (\partial_i)^\mu \\
	\df\Omega & = d\phi \wedge \df n 
	\end{split}\label{eq:NR-choice1}\\[20pt]
\begin{split}
	\tilde{\df e}^0 \NR dt = \df n &\eqsp \tilde{\df e}^a \NR \beta^a_i dx^i \\
	e_0^\mu \NR (\partial_t)^\mu = \tilde v^\mu &\eqsp e_a^\mu \NR \beta_a^i (\partial_i)^\mu \\
	\tilde{\df\Omega} = d\lb(\phi +\half\phi_i\phi^i \rb)&\wedge \df n  + d\phi_i \wedge dx^i
\end{split}\label{eq:NR-choice2}
\end{align}\end{subequations}

Using \autoref{eq:e-omega-trans} and \autoref{eq:coriolis-trans} we can check that these two choices are related precisely by a local boost transformation with \(k_a = \beta^i_a \phi_i\). We could have made more complicated choices for the Lorentzian coframe (with the same Lorentzian spacetime metric) and would have obtained other non-relativistic data all related to each other by a local boost transformation. We think the simple two choices made above illustrate the general principle adequately.

Further, we see that a local boost transformation is precisely the freedom to choose the vector field \(v^\mu\), or in the Lorentzian spacetime the choice of a time-vector field.

Also, in the non-relativistic limit the Newton-Coriolis form encodes the Newtonian potential and effects of choosing a non-inertial frame, and our connection \(\nabla\) through the Christoffel symbols in \autoref{eq:Christ} matches the form in Prop.4.5.2 of \cite{Mal-book} for a Newtonian spacetime. Thus, we can obtain preciely a Newtonian spacetime in the sense of \autoref{sec:barg-restrictions} from the non-relativistic limit of a weak-field Lorentzian spacetime, furthering our confidence in our procedure and definitions. This also emphasizes the role of the boost connection and the associated Newton-Coriolis form as encoding the precise data needed for Newtonian gravity.

From the metric decomposition \autoref{eq:ADM} and choice of $c$-scaling we can also compute the form of the Christoffel symbols,
\be
	{\Gamma^\lambda}_{\mu \nu} \NR  v^\lambda \partial_{(\mu} n_{\nu)} + \frac{1}{2} h^{\lambda \rho} \left( \partial_\mu h_{\nu \rho} + \partial_\nu h_{\mu \rho} - \partial_\rho h_{\mu \nu} \right) + n_{(\mu} \Omega_{\nu )}{}^\lambda,
\ee
where $\df \Omega$ is given by \autoref{eq:Coriolis-NR}. These Christoffel symbols are finite and equivalent to \autoref{eq:Christ} without torsion. Note that this nonrelativistic limit does not require a tensorial redefinition to obtain a sensible limit, unlike the computation of \cite{Jensen:2014wha}.\\

Now in the non-relativistic limit we have seen that \(\df e^0 \NR \df n\) and we introduce the next order (in \(c\)) correction to \(\df e^0\) as
\be\label{eq:n-corr}
	\df e^0 = \df n - \tfrac{1}{c^2}\df a + O(c^{-3})
\ee 
Here, with some foresight, we have already introduced the 1-form \(\df a\), which corresponds to the decomposition of the Poincar\'e time-translation generator in \autoref{eq:Poin-gen}. In the following sections we'll show that it arises precisely as a gauge field for mass in the massive particle and Schr\"odinger Lagrangian in the corresponding non-relativistic limits. Thus, the mass gauge field arises as
\be\label{eq:a-NR}
	-c^2(\df e^0 - \df n)  \NR \df a 
\ee

For the choice of coframe in \autoref{eq:lor-choice} we can compute the mass gauge field as
\begin{subequations}\label{eq:a-NR-choice}\begin{align}
	\df a & = \phi \df n \label{eq:a-NR-choice1}\\
	\tilde{\df a} &= \lb(\phi + \half\phi_i\phi^i\rb)\df n + \phi_i dx^i \label{eq:a-NR-choice2}
\end{align}\end{subequations}
which are also related by a boost with \(k_a = \beta_a^i\phi_i\) as seen from \autoref{eq:e-omega-trans}.\\

Also, we see that non-relativistic limits of torsionless Lorentzian spacetimes always yield a Newtonian spacetime. In particular the mass torsion always vanishes
\be
	-d \lb(c^2(\df e^0 - \df n )\rb) + c^2d\df e^0 \NR d\df a - \df \Omega = \df f = 0 
\ee\\

We can account for torsion in the spacetime, without much difficulty. In fact, one can identify clearly the origins of \(\df f\) in the non-relativistic limit. Using a torsionful Lorentzian spacetime \autoref{eq:a-NR} does not change while \autoref{eq:Coriolis-NR} does get a contribution from \(\df T^0\) and we see that
\be
	-c^2\df T^0 \NR \df f
\ee
Thus, a non-relativistic limit of torsionful Lorentzian spacetimes where \(\df T^0 = O(c^{-2})\) will necessarily give a non-zero mass torsion. One must be careful in case the torsion is \(\df T^0 = O(1)\), since we can not identify \(\df n = dt\) without violating the Newton-Cartan condition \(\nabla_\mu n_\nu = 0\) in the non-relativistic limit. In this case, we can choose the ADM lapse to scale as \(N^2 = e^{-2\Phi_L}\lb(1 - \frac{2\phi}{c^2}\rb) + O(c^{-3})\), where the Luttinger potential is \(\Phi_L = O(1)\), and the torsion to be \(\df T^0 = -e^{-\Phi_L} ~d\Phi_L \wedge dt + O(c^{-2})\) which gives us a causal Bargmann spacetime in the non-relativistic limit.\\

Thus, we can get all causal Bargmann spacetimes as the non-relativistic \(c \to \infty\) limits of appropriate Lorentzian spacetimes.

%%------------------------------------------------------

\subsection{Massive particle from Lorentzian worldline}

Consider a time-like wordline \(\gamma\) in the Lorentzian spacetime with tangent \(T^\mu\). The proper time measured along \(\gamma\) is then (see \cite{Wald-book})
\be
	\tau = \int_\gamma d\lambda~ (-T^\mu T_\mu)^\half
\ee

For the non-relativistic limit we compute
\be\begin{split}
	-T^\mu T_\mu &= -g_{\mu\nu}T^\mu T^\nu = -\eta_{AB}e^A_\mu e^B_\nu T^\mu T^\nu \\
		& = \lb( c^2 e^0_\mu e^0_\nu - \delta_{ab}e^a_\mu e^b_\nu \rb)T^\mu T^\nu \\
		& = \lb( c^2n_\mu n_\nu - G_{\mu\nu} \rb)T^\mu T^\nu
\end{split}\ee
where we have used the form \autoref{eq:n-corr} for the Lorentzian coframe and the extended metric from \autoref{eq:ext-G}. We see that
\be
	c^{-1}\tau \NR \int_\gamma d\lambda~ T^\mu n_\mu = \int_\gamma \df n
\ee
The \(c\)-scaling of proper time above, is exactly what is expected from simple consideration of units. Similarly, for the corresponding non-relativistic curve we take the tangent vector to be \(cT^\mu \NR T^\mu\) and thus the normalized tangent behaves as \(\xi^\mu \NR \xi^\mu\). In the Lorentzian spacetime the action for geodesics is given by
\be
	\mc S_{\rm particle} = -\half \int_\gamma d\tau~ g_{\mu\nu}\xi^\mu\xi^\nu \NR -\half \int_\gamma d\tau~ \lb( c^2 n_\mu n_\nu - G_{\mu\nu} \rb)\xi^\mu\xi^\nu
\ee
The first term is a constant that can be safely ignored from the action and thus we get
\be
	\mc S_{\rm particle} \NR \half \int_\gamma d\tau~ G_{\mu\nu}\xi^\mu\xi^\nu = \frac{1}{2}\int_\gamma d\tau~ {\vd h}_{\mu\nu}\xi^\mu \xi^\nu + \int_\gamma \df a
\ee
which matches the action \autoref{eq:particle-action-NC}. The above computation shows that the non-relativistic particle on any causal Bargmann spacetime can be obtained as a limit of Lorentzian time-like geodesics.\\

%--------------------------------------------------------
\subsection{Schr\"odinger from Klein-Gordon}

Next we consider the non-relativistic limit of the massive Klein-Gordon field to a massive Schr\"odinger field \cite{Kuchar-Sch, DK, DBKP}. On the Lorentzian spacetime consider a complex scalar field \(\Phi\) of mass \(m\) with dynamics given by the Klein-Gordon Lagrangian
\be\label{eq:KG-Lag}
	\mc L_{KG} = -g^{\mu\nu}\nabla_\mu\bar\Phi\nabla_\nu\Phi - m^2c^2\bar\Phi\Phi
\ee

Given a time function \(t\), we identify the corresponding Schr\"odinger field \(\psi\) as
\be\label{eq:KG-Sch}
	\Phi = e^{-imc^2t}\psi
\ee
which gives
\be\label{eq:KG-NR-Sch}
	\nabla_\mu\Phi = e^{-imc^2t}\lb( -imc^2n_\mu + \nabla_\mu \rb)\psi
\ee
where we have used the identification \(\df n = dt\), which in the limit, corresponds to restricting to Bargmann spacetimes with absolute time (\(d\df n = 0\)). We further note that \(e_A^\mu e^0_\mu = \begin{pmatrix}1,0\end{pmatrix}\) and hence using \autoref{eq:n-corr}
\be\label{eq:NR-e-id}\begin{split}
	e_A^\mu n_\mu & = \begin{pmatrix}1,0\end{pmatrix} + \tfrac{1}{c^2}a_A + O(c^{-3}) \\
	-g^{\mu\nu}n_\mu n_\nu & \approx \frac{1}{c^2}\lb(1 +\frac{a_0}{c^2}\rb)^2 - \frac{1}{c^4}a_a a^a \approx \frac{1}{c^2} + \frac{2a_0 - a_aa^a}{c^4} 
\end{split}\ee

Now we are in a position to take the non-relativistic limit of the Klein-Gordon Lagrangian \autoref{eq:KG-Lag}. In what follows we'll denote \(\nabla_A \defn e_A^\mu \nabla_\mu \) for brevity
\be\begin{split}
	\mc L_{KG} & = - \eta^{AB}\nabla_A\bar\Phi\nabla_B\Phi - m^2c^2\bar\Phi\Phi \\
		& = - \eta^{AB}\lb(imc^2e_A^\mu n_\mu\bar\psi + \nabla_A\bar\psi\rb) \lb(-imc^2e_B^\nu n_\nu\psi + \nabla_B\psi\rb) - m^2c^2\bar\psi\psi
\end{split}\ee

Expanding this and using the identities in \autoref{eq:NR-e-id} gives
\be\label{eq:KG-Sch-Lag}\begin{split}
	\mc L_{KG} & \NR \bar\psi m^2 (2a_0 - a_a a^a)\psi + im\bar\psi \nabla_0\psi + im\bar\psi a^a \nabla_a\psi \\
		&\qquad - im\nabla_0\bar\psi \psi - im\nabla_a\bar\psi a^a \psi - \nabla_a\bar\psi \nabla^a\psi \\
		& = im \bar\psi e_0^\mu \lb( \nabla_\mu\psi - im a_\mu \psi \rb) -im e_0^\mu \lb( \nabla_\mu\bar\psi + im a_\mu \bar\psi \rb)\psi \\
		&\quad - h^{\mu\nu}\lb( \nabla_\mu\bar\psi + im a_\mu \bar\psi \rb) \lb( \nabla_\nu\psi - im a_\nu \psi \rb) \\
		& =  im \bar\psi v^\mu D_\mu\psi - im v^\mu D_\mu\bar\psi \psi  - h^{\mu\nu}D_\mu\bar\psi D_\nu\psi
\end{split}\ee
where \(D_\mu = \nabla_\mu - a_\mu \gen M\) is the covariant derivative for massive fields with \({\gen M} \psi = im\psi\) and \({\gen M} \bar\psi = -im\bar\psi\). Thus, we retrieve the covariant Lagrangian for a Schr\"odinger field \autoref{eq:Sch-action-NC} as a non-relativistic limit of the Klein-Gordon Lagrangian.\\

We note that on a causal Bargmann spacetime with \(d\df n \neq 0 \) the ``standard" identification between the Klein-Gordon and Schr\"odinger fields in \autoref{eq:KG-Sch} does not give a well-defined \(c \to \infty\) limit for the Klein-Gordon Lagrangian\footnote{We thank Andreas Karch and Kristan Jensen for pointing this out to us.}. As long as the spacetime is causal we still have a notion of global time (See \autoref{sec:barg-restrictions}) which we may use in our identification, but an identification of the form $\Phi=e^{-imc^2 f(x^\mu)}\psi$ can no longer simultaneously cancel both $O(c^4)$ and $O(c^2)$ terms. Even though we can still write the Schr\"odinger Lagrangian \autoref{eq:Sch-action-NC} \emph{directly} on such a Bargmann spacetime, it would be of interest to figure out the correct identification to obtain it from a non-relativistic limit.\\

The above computations show the importance of \(\df\varpi^a\), \(\df a\) and \(\df f\) for coupling massive fields to a Bargmann spacetime. They capture the Lorentzian data that is next-to-leading order in \(c\) while taking the non-relativistic limit. This includes effects of torsion, non-inertial frames, and most importantly, Newtonian gravity.

%%=========================================================

\section{Outlook}\label{sec:outlook}

In this paper we have constructed the most general geometric backgrounds consistent with local Galilean invariance, in particular paying attention to the role that played by torsion. We have identified the freedom of the connection not fixed by compatibility with $\df n$ and $h^{\mu\nu}$ and spatial torsion. There are a number of further directions worth investigating.

It would also be very useful to understand how to classify the symmetries of a Bargmann spacetime. Symmetries of Lorentzian spacetimes, are diffeomorphisms (generated by a vector field \(\xi\)) which preserve the metric and the torsion, $\pounds_\xi g_{\mu \nu} = \pounds_\xi T^\lambda{}_{\mu\nu}=0$. In Newton-Cartan spacetimes it is not enough to require $\pounds_\xi n_\mu = \pounds_\xi h^{\mu\nu} = \pounds_\xi T^\lambda{}_{\mu\nu} = 0$, as there is still data in the connection that is undetermined by the clock form, metric and torsion. Understanding the correct definition of symmetries is a crucial step in understanding how to classify geometries. This would most easily be done by studying the automorphisms of the principal \(Barg(1,d)\) bundle (see \cite{Chamel:2014fta} for some work in this direction.).\\

Another interesting problem is to understand how to write actions for Newtonian gravity. The field equation for Newtonian gravity is very well known:
\be
R_{\mu\nu} = 4 \pi G_N \rho~ n_\mu n_\nu
\ee
 Though it is surprisingly difficult to write down an action for Newtonian gravity in terms of spacetime tensors. \cite{Goenner} introduces a Lagrangian in terms of certain matter field variables and Lagrange multipliers while \cite{DBKP} prescribes a Lagrangian in a Lorentzian spacetime and the field equations are then derived via null reduction. As pointed out earlier, the gravitational Lagrangian in \cite{Jensen:2014aia} does not give Newtonian gravity. It would be interesting to see if the extended frame formalism introduced in this work can be used to write a manifestly invariant Lagrangian for Newtonian gravity.\\

There is also a curious relation between our extended formalism and the \emph{tractor calculus} used in the study of conformal geometry\cite{CRG}. For instance, the action of the Galilean group in the extended representation \autoref{eq:gal-group-ext} matches the action of conformal transformations on sections of the tractor bundle Eq.26 of \cite{CRG}. Also, the extended metric in \autoref{eq:inv-tensors} and \autoref{eq:ext-G} are the same as the \emph{tractor metric} in Prop.3.12 in \cite{CRG}. We think that these hint at some interesting relationships between the Bargmann geometry studied in this work and conformal geometry that deserve more exploration.

%%=========================================================

\section*{Acknowledgments}
We would like to thank Robert M. Wald for numerous discussions and for pointing out the extensive relativity literature on Newton-Cartan geometry. We also thank  Dam T. Son for insightful discussions and constant encouragement.  M.G. is supported in part by NSF grant DMR-MRSEC 1420709. K.P. is supported in part by NSF grant PHY 12-02718 to the University of Chicago. M.M.R is supported in part by DOE grant DE-FG02-13ER41958.

\bibliographystyle{JHEP}
\bibliography{bargmann_geometry}

%%%%%%%%%%%%%%%%%%%%%%%%%%%%%%%%%%%%%%%%%%%%%%%%%%%%%%%%%%%%%%%%%
\end{document}